\documentclass[twocolumn, astrosymb]{aastex631}

\usepackage{amsmath}
\usepackage[perpage]{footmisc} 

\usepackage{macros}
\usepackage{color}

\shortauthors{Farah, Edelman, et al.}

\graphicspath{{./}{figures/}}

\begin{document}

\title{Things that might go bump in the night: Assessing structure in the binary black hole mass spectrum}

\author[0000-0002-6121-0285]{Amanda M. Farah}
\email{afarah@uchicago.edu}
\affiliation{Department of Physics, University of Chicago, Chicago, IL 60637, USA}

\author[0000-0001-7648-1689]{Bruce Edelman}
\email{bedelman@uoregon.edu}
\affiliation{Institute  for  Fundamental  Science, Department of Physics, University of Oregon, Eugene, OR 97403, USA}

\author[0000-0002-0147-0835]{Michael Zevin}
\affiliation{Kavli Institute for Cosmological Physics, The University of Chicago, 5640 South Ellis Avenue, Chicago, Illinois 60637, USA}
\affiliation{Enrico Fermi Institute, The University of Chicago, 933 East 56th Street, Chicago, Illinois 60637, USA}

\author[0000-0002-1980-5293]{Maya Fishbach}
\affiliation{Canadian Institute for Theoretical Astrophysics, David A. Dunlap Department of
Astronomy and Astrophysics, and Department of Physics, 60 St George St, University of Toronto, Toronto, ON M5S 3H8, Canada}

\author[0000-0002-7213-3211]{Jose María Ezquiaga}
\affiliation{Niels Bohr International Academy, Niels Bohr Institute, Blegdamsvej 17, DK-2100 Copenhagen, Denmark}

\author[0000-0002-2916-9200]{Ben Farr}
\affiliation{Institute  for  Fundamental  Science, Department of Physics, University of Oregon, Eugene, OR 97403, USA}

\author[0000-0002-0175-5064]{Daniel E. Holz}
\affiliation{Department of Physics, University of Chicago, Chicago, IL 60637, USA}
\affiliation{Kavli Institute for Cosmological Physics, The University of Chicago, 5640 South Ellis Avenue, Chicago, Illinois 60637, USA}
\affiliation{Enrico Fermi Institute, The University of Chicago, 933 East 56th Street, Chicago, Illinois 60637, USA}

\begin{abstract} 
Several features in the mass spectrum of merging binary black holes (BBHs) have been identified using data from the Third Gravitational Wave Transient Catalog (GWTC-3). 
These features are of particular interest as they may encode the uncertain mechanism of BBH formation.
We assess if the features are statistically significant or the result of Poisson noise due to the finite number of observed events. 
We simulate catalogs of BBHs whose underlying distribution does not have the features of interest, apply the analysis previously performed on GWTC-3, and determine how often such features are spuriously found.
We find that \new{one} of the features found in GWTC-3, the peak at $\sim35\Msun$, cannot be explained by Poisson noise alone: peaks as significant occur in \result{$1.7\%$} of catalogs generated from a featureless population.
This peak is therefore likely to be of astrophysical origin.
\new{The data is suggestive of an additional significant peak at $\sim10\Msun$, though the exact location of this feature is not resolvable with current observations.}
Additional structure beyond a power law, such as the purported dip at $\sim14\Msun$, can be explained by Poisson noise.
We also provide a publicly-available package, \texttt{GWMockCat}, that creates simulated catalogs of BBH events with \new{correlated} measurement uncertainty and selection effects according to user-specified underlying distributions and detector sensitivities.
\end{abstract}

\section{Introduction}
\label{sec:intro}
Gravitational waves (GWs) from more than 70 mergers of compact objects have now been detected in the data of the LIGO \citep{LIGO_det} and Virgo \citep{Virgo_det} detectors.
A cumulative catalog of these events and their properties has been produced by the LIGO-Virgo-KAGRA (LVK) collaborations.
This collection of all detections to date is called the ``Third Gravitational-Wave Transient Catalog'' \citep[GWTC-3]{gwtc3}, and has enabled several insights into the nature of gravity~\citep{gwtc3_TGR} \comment{(others?)}, the local expansion of the universe~\citep{gwtc3_cosmo}\comment{(others?)}, and the population of GW sources \citep{O3b_pop}\comment{(others?)}.

The underlying population of GW sources holds information about the astrophysical processes that give rise to merging binaries of compact objects.
The mass spectrum of binary black holes (BBHs), for example, encodes information about numerous physical processes underlying massive-star evolution, supernova physics, compact object formation, and binary interactions. 
For example, the presence or dearth of black holes with masses between $\sim 2$--$5~\Msun$~\citep{2010ApJ...725.1918O, 2011ApJ...741..103F, fishbach_does_2020, farah_bridging_2022}\comment{(others?)} may unveil the maximum neutron star mass, the stability of mass transfer, and the timescales relevant for the engines that drive supernova explosions~\citep[e.g.,][]{fryer_compact_2012,zevin_exploring_2020,mandel_simple_2020,li_constraints_2021,van_son_no_2022,patton_comparing_2022,siegel_investigating_2022}. 
On the high mass end, a sharp decrease in the mass spectrum for black holes with masses $\gtrsim 50~\Msun$~\citep{2017ApJ...851L..25F,2021ApJ...913L..23E} would be a strong indication that the pair instability process is at play and limiting the core mass of massive stars~\citep{fowler_neutrino_1964,barkat_dynamics_1967,heger_nucleosynthetic_2002,heger_how_2003,woosley_deaths_2015,2016A&A...594A..97B, woosley_pulsational_2017,woosley_evolution_2019,marchant_pulsational_2019,renzo_predictions_2020}, with the location of the decrease in the differential merger rate acting to constrain relevant nuclear reaction rates~\citep{farmer_constraints_2020}. 
Other overdensities and underdensities in the observed mass distribution \citep{Edelman_2022,Tiwari_2021,tiwari_exploring_2022,edelman_cover_2022}, as well as the evolution of the mass distribution with redshift \citep{fishbach_when_2021,van_son_redshift_2022,karathanasis_binary_2022,van_son_no_2022}, will further inform the dominant BBH formation channels, binary evolution physics, and the metallicity evolution of the universe. 

All of the parameters that are measurable from the signal of a binary merger can provide insight into formation mechanisms of merging binaries, especially when used in a population analysis \citep{stevenson_distinguishing_2015, zevin_constraining_2017}.
However, the masses of the objects in the merging system are the best measured and span the largest dynamic range.
Additionally, the mass distribution of compact objects can be used to measure cosmological parameters using the ``spectral siren'' method, provided there is structure in the distribution beyond a boundless power law \citep{chernoff_gravitational_1993,messenger_measuring_2012, 2012PhRvD..85b3535T,farr_future_2019, 2021ApJ...909L..23E,ezquiaga_spectral_2022, gwtc3_cosmo}, such as edges, gaps, peaks, or changes in the power law slope.
Multiple features must be present to disentangle redshift evolution of the mass spectrum from cosmology, and more features further aid in breaking this degeneracy \citep{ezquiaga_spectral_2022}.
Therefore, considerable effort in the field of GW astronomy has gone towards understanding the mass distribution of GW sources.
There are currently many more detected BBH mergers than binary neutron star (BNS) or neutron star-black hole (NSBH) mergers, so much of the activity has been on population properties of the BBH distribution, though the mass distribution of BNSs and NSBHs has also been been considered \citep{fishbach_does_2020, landry_mass_2021, farah_bridging_2022, ye_inferring_2022, biscoveanu_population_2022}.

\begin{figure}
    \centering
    \includegraphics[width=\columnwidth]{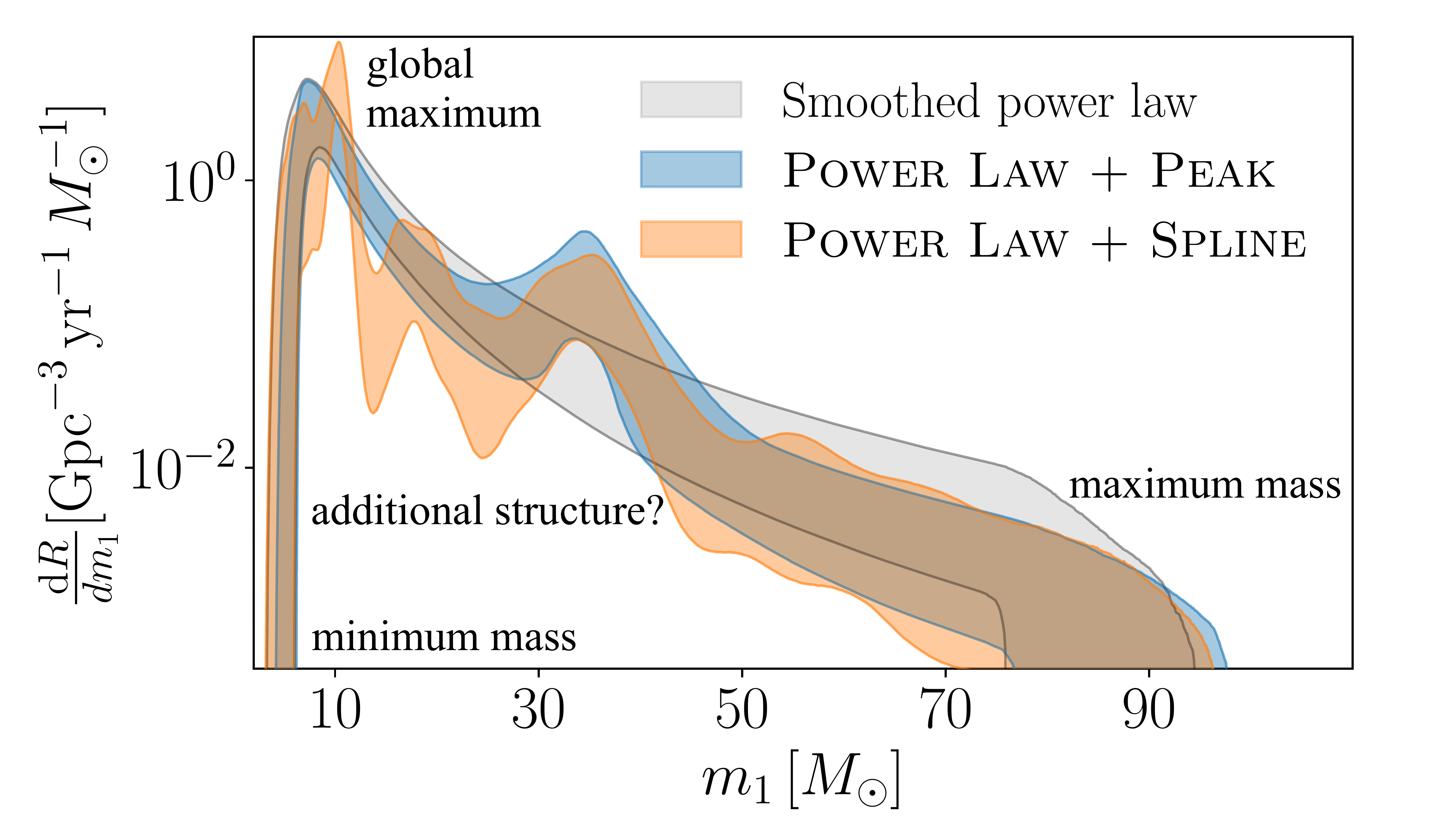}
    \caption{Distribution of primary BBH masses inferred using GWTC-3 and three different population models.
    The smoothed power law model (\emph{grey}) consists of a single power law slope between a minimum and maximum mass, with the merger rate set to exactly zero outside of those bounds.
    It also includes a smoothing parameter at the low-mass end that allows for an offset between the minimum BH mass and the global maximum of the distribution.
    The \textsc{Power Law + Peak} model is similar to the smoothed power law, but also includes a Gaussian component.
    The \PS{} model adds a cubic spline modulation to a smoothed power law to allow for additional substructure.
    We seek to determine if the perturbations beyond a power law found by \PS{} and other semi-parametric models can be explained by random associations in the data due to a finite number of observations, or if they are features of the true underlying distribution.
    }
    \label{fig:dR_dm1}
\end{figure}

The \new{mass distribution of merging BBHs} is typically parameterized by the primary mass $m_1$, the larger of the two component masses in the binary, and the mass ratio $q=m_2/m_1$, the ratio of the less massive object's mass to the primary mass, though other parameterizations are possible and valid \citep[e.g.,][]{farah_bridging_2022, fishbach_picky_2020, Tiwari_2021}.
The community has thus far gained a robust understanding of the large-scale features of the BBH mass distribution, and is just beginning to resolve its finer details.
After the release of the First Gravitational-Wave Transient Catalog \citep[GWTC-1]{gwtc1}, minimum and maximum masses at $\sim5\,\Msun$ and $\sim40\,\Msun$ were identified in the BBH primary mass distribution, but it was not yet possible to distinguish between a uniform distribution and a power law between those two bounds \citep{2017ApJ...851L..25F,talbot_measuring_2018,GWTC1_pop}.
The Second Gravitational-Wave Transient Catalog \citep[GWTC-2]{gwtc2} brought dozens of additional events, and the BBH mass distribution was found to have 
a global maximum at $\sim 8\Msun$ and an excess of BHs between $\sim 30\Msun$--$40 \Msun$ followed by a steep, although not infinitely sharp, drop off in the rate at higher masses extending to $\sim80\,\Msun$ (instead of sharp cutoff at $\sim40\,\Msun$).
At the time, there were not enough observations to determine whether the mass distribution had a local maximum at $\sim35\,\Msun$, represented by a Gaussian peak on top of a power law, or whether the steepening towards higher masses was better described as a break in the power law~\citep{O3a_pop}.

At the end of the third LIGO--Virgo observing run, the same two features at $\sim8\Msun$ and $\sim35\,\Msun$ remained, and the feature at $35 \Msun$ was classified as a peak rather than a break in the power law \citep{O3b_pop}.
Additionally, non-parametric \citep{mandel_model-independent_2017,rinaldi_hdpgmm_2022,Sadiq_2022,payne_np_22,edelman_cover_2022} and semi-parametric \citep{Edelman_2022} analyses found robust evidence for an additional peak at $\sim10\Msun$, the same peak at $\sim35\Msun$, as well as modest evidence for a paucity of events near $\sim 14\Msun$ \citep{O3b_pop}.
These features in the primary mass distribution correspond to similar ones in the chirp mass distribution, occurring at $\sim 9 \Msun$, $\sim 11 \Msun$, and $\sim 26 \Msun$, respectively \citep{Tiwari_2021,tiwari_exploring_2022}.
The current picture of the BBH mass distribution is therefore a decreasing power law from low to high masses, with a global maximum at $m_1 \sim 10 \Msun$, a potential underdensity at $m_1 \sim 14 \Msun$, and an overdensity at $m_1\sim 35 \Msun$. 
This can be see\new{n} in Figure~\ref{fig:dR_dm1}, where we plot the results of fitting two parameteric models and one semi-parametric model to the BBHs in GWTC-3.

While the existence of this substructure in the current data set appears robust, its interpretation is less clear.
Plausible explanations for this substructure include (1) Poisson noise, (2) modeling systematics, or (3) astrophysical signatures from one or several formation channels.
We aim to disentangle the first two possibilities from the third using the \PS{} model \citep{Edelman_2022}, one of the semi-parametric models used to identify the substructure reported in \citet{O3b_pop}.

Poisson noise would be caused by the fact that the fiducial BBH analysis in \cite{O3b_pop} includes only 69 events over a mass range that spans more than an order of magnitude, so observations may appear to be clumped at some masses even if the underlying distribution is smooth.
We first determine if this explanation accounts for the data by simulating catalogs of BBHs whose underlying distribution does not have the features of interest, applying the analysis previously performed on GWTC-3, and determining how often such features are spuriously found.
We develop several metrics comparing observations to simulated data in order to assess the statistical significance of the ``bumps'' in the primary mass distribution found by \cite{O3b_pop,Edelman_2022}.
All of the metrics derived in this work answer the same general question: how often do we infer the existence of a feature when analyzing observations of a true population \emph{without} that feature?
In this sense, these metrics are analogous to frequentist $p$-values, as lower values correspond to more significant features in the data.
Readers familiar with gravitational wave data analysis might find it useful to think of these metrics as false alarm rates because they quantify how often noise resembles the observed signal.

A similar frequentist analysis on a large number of mock catalogs was performed by \cite{Sadiq_2022} on the peak at $\sim35\Msun$ using an adaptive kernel density estimator (aKDE) to find features in samples drawn from featureless mass models, as well as from a model with a single peak.
They account for selection effects, but not measurement uncertainty.
They find that an aKDE is able to identify peaks in the data, and that the peak at $\sim35\Msun$ found in GWTC-2 is statistically significant within the aKDE model.

The second effect mentioned above, model systematics, could also plausibly cause spurious inference of features beyond a power law.
It is potentially concerning that the models considered in \cite{O3b_pop} that find peaks and troughs in the mass distribution are inherently ``bumpy'': both \textsc{Power Law + Peak} \citep{talbot_measuring_2018} and \textsc{Multi source} employ a smoothed power law with a Gaussian component \citep{wysocki_popmodels_2021}, \textsc{Flexible Mixtures} is a linear combination of Gaussian components, and \PS{} employs a smoothed power law under a cubic spline modulation.
The question is then whether these ``bumpy'' models can recover sharp features or if they instead create peaks and troughs that are morphologically dissimilar to the true distribution. 
This is most easily addressed by cross-checking with independent models such as \textsc{Broken Power Law} \citep{O3a_pop, O3b_pop} and the auto-regressive model presented in \cite{callister_parameter-free_2023}.

Inaccuracies in the selection function are also known to cause systematic biases when inferring the underlying population \citep[e.g.][]{malmquist_relations_1922,malmquist_contribution_1925}.
These biases could, in principle, also cause an incorrect inference of structure in the astrophysical distribution of BBH masses.
However, selection effects in GW detectors are remarkably well-characterized, so we expect this effect to be subdominant to Poisson uncertainty.
As the number of events grows, so will our accuracy in the estimation of the selection function~\citep{farr_accuracy_2019,essick_precision_2022}.

We provide posterior samples from our simulated catalogs in an accompanying data release \citep{farah_data_2022}, and also provide a publicly-available \texttt{python} package, \texttt{GWMockCat} \citep{farah_gwmockcat_2022}, to create similar samples according to user-defined populations.\footnote[1]{The data release can be found at \dataset[DOI: 10.5281/zenodo.7411991]{https://doi.org/10.5281/zenodo.7411991}, and \texttt{GWMockCat} can be installed at \url{https://git.ligo.org/amanda.farah/mock-PE} .}

Section~\ref{sec:results-1} provides a demonstrative example: it foregoes a full fit to the astrophysical population of sources, and compares the observed distribution of masses to possible observed distributions given an underlying power law in primary mass, (incorrectly) assuming no measurement uncertainty.
This analysis suggests that the observed peak at $\sim35\Msun$ is statistically significant, but that all other features beyond a simple power law might be explainable by Poisson noise.
This motivates a thorough study using a full hierarchical Bayesian analysis on simulated event posteriors, which we carry out in Section~\ref{sec:full}. Section~\ref{sec:discussion} summarizes our conclusions and discusses their implications for the astrophysical origin of the gravitational waves observed thus far by the LVK.
Readers primarily interested in the significance of features in the mass distribution may wish to skip to Section~\ref{sec:results-2}, whereas those interested in using the package \texttt{GWMockCat} can find details in Appendices~\ref{appendix:mockPE} and~\ref{appendix:MDC}.

\section{Motivation} 
\label{sec:results-1}
\begin{figure}
    \centering
    \includegraphics[width=\columnwidth]{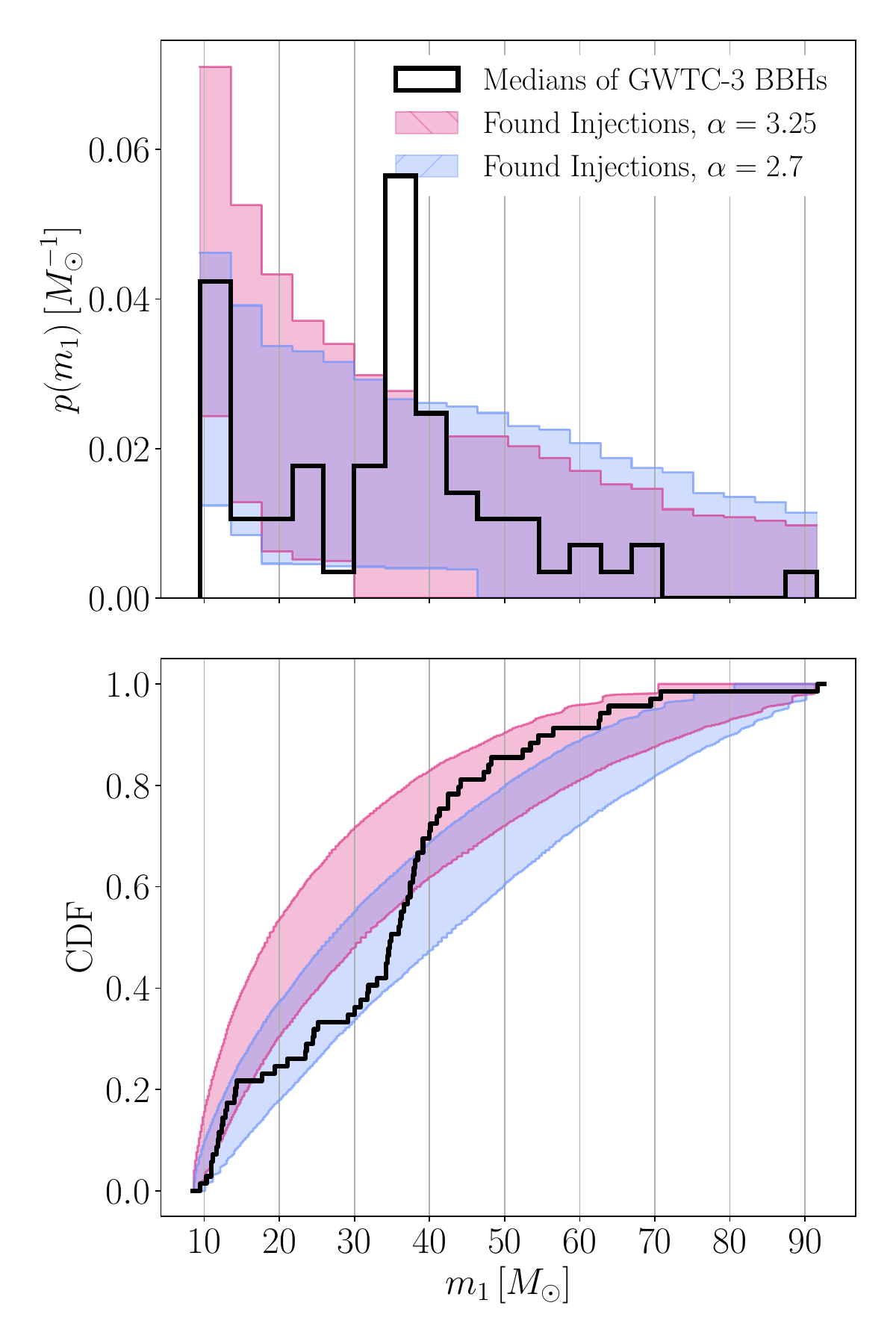}
    \caption{Observed source-frame primary mass distributions.
    Black solid lines contain the median \emph{a posteriori} values for the binary black holes in GWTC-3.
    Pink and blue bands indicate the 90\% credible interval on the \emph{observed} distributions predicted from astrophysical distributions that are power laws in primary mass with spectral index $\alpha=3.25$ and $\alpha=2.7$, respectively.
    The top panel shows a histogram of observed primary masses.
    For GWTC-3's distribution to be consistent with the null distributions, we expect its bin heights in the top panel to be within the 90\% credible intervals in 18 out of the \result{20} bins.
    The uncertainties in these predicted distributions are due only to Poisson noise resulting from a finite number of observations, rather than modeling uncertainty or uncertainty in parameter estimation.
    Therefore, the cumulative distribution functions in the bottom panel are similar to a conventional posterior predictive check, but with only one source of uncertainty.
    The large deviations of the black curve from the shaded bands in some regions indicate the difficulty that a single power law with  Poisson shot noise has in fully explaining the observations.
    However, many of the apparent excursions from a power law are well-contained within the predicted bands.
    }
    \label{fig:injection_CDF}
\end{figure}

To construct a simple test of feature significance and motivate further study, we first avoid a fit to the mass distribution and instead consider the \emph{observed} distribution of primary masses and its resemblance to one that would result from a simple power law.
The observed population differs significantly from the astrophysical one, as current gravitational wave detectors are subject to selection biases that favor the detection of closer and more massive systems, as well as measurement error that affects each system differently.
We construct plausible observed mass distributions that could occur from detecting 69 BBHs whose astrophysical distribution is a featureless power law in primary mass.
To do this, we use the samples provided by \cite{O3_injections_data_release}, which were created for sensitivity estimation for the LVK's GWTC-3 analysis.
Each of these samples comes with a probability of being drawn from an assumed underlying distribution and a false alarm rate (FAR) assigned by each search used by the LVK.
We can then re-weight these samples to our desired population model (in this case, a power law in $m_1, q$, and $z$) using the draw probability, and apply the same FAR threshold used in \cite{O3b_pop} to select ``found injections.''
Of the \result{$\sim 6 \times 10^{4}$} found injections, we resample to \result{$N = 10^{4}$} independent sets of \result{69} draws each to directly compare to observations.

We then histogram each set of these found injections, thereby obtaining a distribution of bin heights for our mock populations.
Using several thousand realizations of found injection sets enables us to construct a null distribution of bin heights and characterize the effect of Poisson noise on the shape of the observed distribution.
We compare these null distributions with the observed distribution of BBH masses in GWTC-3\footnote[2]{For all comparisons to real observations, we use the publicly available posterior samples for the GWTC-2.1 and GWTC-3 data releases \citep[][respectively]{gwtc2p1_data_release, gwtc3_data_release}.
We use samples generated with the IMRPhenomXPHM waveform and a prior proportional to the square of the luminosity distances (i.e. the samples were not ``cosmologically reweighted'').
To make the most direct comparison with \cite{O3b_pop}, we keep events with secondary mass larger than $3\Msun$ and FAR less than $1\, \text{yr}^{-1}$, resulting in 69 events.}
by assuming the primary masses are measured perfectly and using the median \emph{a posteriori} values of their primary masses as point estimates.
The result is shown in the top panel of Figure~\ref{fig:injection_CDF}, which plots the 90\% credible interval on the observed null distributions, along with the distribution of median primary masses of GWTC-3's BBHs.
For the null distributions, we consider two power law spectral indices as representative examples: $\alpha=2.7$ and $\alpha=3.25$.
These are chosen to represent a range of plausible values for the BBHs in GWTC-3: a power law fit to GWTC-3 yields $\alpha=2.98^{+0.16}_{-0.28}$, where the bounds represent 1-$\sigma$ deviations.

To obtain a more quantitative measure, we compare bin heights from the found injections, $h_{\text{inj}}$, to the bin heights of observed events in GWTC-3, $h_{\text{GWTC-3}}$, obtaining for each bin $i$ the fraction of simulated bin heights that are lower than those of GWTC-3 BBHs.
Explicitly,
\begin{equation}
    r^i_h = \frac{1}{N}\sum^N_j \begin{cases}
        1 &\text{ if } h_{i,\text{inj}}^j < h_{i,\text{GWTC-3}} \\
        0 &\text{ if } h_{i,\text{inj}}^j \geq h_{i,\text{GWTC-3}}
    \end{cases} ,
    \label{eq:rhi}
\end{equation}
where the sum is over the \result{$N = 10^{4}$} sets of found injections, and $r_h$ is defined for each bin. 
For example, if $r_h=0.95$ for a given bin, the observed distribution in that bin is larger than would be expected from a featureless power law $95\%$ of the time.
A value of $r_h$ approaching unity corresponds to a ``bump'' in the observed mass distribution, and a value of $r_h$ approaching zero is indicative of a ``dip.''

Note that the comparison between the null distributions and GWTC-3 are occurring at each bin, rather than across all bins.
We do this because the magnitude of Poisson noise depends on the value of $m_1$: since the underlying distribution is not uniform, fewer events are expected at very high $m_1$ and therefore the relative standard deviation is larger.
This is also a consequence of Eddington bias \citep{1913MNRAS..73..359E}.
Making comparisons at specific points in $m_1$ does not, however, properly correct for the look-elsewhere effect.
We will address this effect in Section~\ref{sec:full}.

The three most significant values of $r_h^i$ in the case of $\alpha=3.25$ are \result{$r_h^{15.6\Msun}=0.033$, $r_h^{27.9\Msun}=0.036$, $r_h^{36.1\Msun}>0.999$}, where the superscripts indicate the centers of the bins at which $r$ was calculated.
This means that \result{less than 0.1\%} of mock populations had more events near \result{$36.1\Msun$} than GWTC-3 does, \result{3.3\%} of mock populations had fewer events near \result{$15.6\Msun$} than GWTC-3, and at \result{$27.9\Msun$, 3.6\%} of mock populations had fewer events.

Repeating the exercise for $\alpha=2.7$, we find the three most significant values of $r_h^i$to be \result{$r_h^{40.2\Msun}=0.935$, $r_h^{27.9\Msun}=0.020$, $r_h^{36.1\Msun}>0.999$}. 
The locations of the significant features differ when the assumed underlying distribution changes\footnote[3]{It is also possible to determine the existence of local minimia or maxima in this observed distribution independently of the underlying power law.
This can be done using a dip test for unbinned data \citep{hartigan_dip_1985} or the minimum number of components required for a Gaussian mixture model \citep{mclachlan_multivariate_2000}.
However, features in the observed distribution would be difficult to disentangle from selection effects, so we recommend only applying these to the astrophysical distribution, as in \citet{Tiwari_2021}.
Since our principal aim is to quantify the significance of excursions from a power law, we leave such tests for future work.}.
In either case, the bump at $\sim35\Msun$ is unlikely to be due to Poisson noise, but other features may be.

To avoid the need to arbitrarily choose bins, we additionally construct a cumulative distribution function (CDF) of the primary masses and compare it to the CDFs of the null distributions, shown in the bottom panel of Figure~\ref{fig:injection_CDF}.
This comparison is akin to a posterior predictive check in that it can highlight where the model fails to predict the data.
Importantly, though, it differs from the conventional posterior predictive check because we have purposefully left out the effects of modeling uncertainty and measurement uncertainty in order to isolate the effects of Poisson noise.
The prior distributions are therefore also not included, since each event is assumed to be measured with perfect accuracy. 

If $\alpha=3.25$, the null distributions are consistent with the data below $\sim18\Msun$ and above $\sim 35\Msun$, but not between them, meaning that the $\sim10\Msun$ and $\sim35\Msun$ peaks can be explained by Poisson noise, but the underdensity between them could not be.
On the other hand, if $\alpha=2.7$, the null distributions are consistent with the data everywhere except for above $\sim40\Msun$, suggesting that under this scenario, Poisson noise can explain all features except for the $\sim35\Msun$ peak.

For both spectral indices considered, two of the three features found by \cite{O3b_pop} can be explained by Poisson noise from a finite number of observations.
However, this does not mean that exactly two of the features are the result of Poisson noise, just that no more than two can be caused by the phenomenon.
Additionally, it is not clear \emph{which} of the features are more likely to have physical origin, as this method offers no quantitative way to determine which power law slope is preferred.

Importantly, this methodology does not account for the effects of measurement error, which can cause significant biases near the edges of sharp distributions when not properly accounted for~\citep{FFH}.
We therefore turn to a full hierarchical Bayesian analysis of simulated catalogs, which will allow us to fit for population model parameters, take measurement uncertainty into account, and directly compare to metrics used in \cite{O3b_pop}.



\section{Hierarchical analysis and Results}
\label{sec:full}
We determine how often the features inferred in the mass distribution of BBHs would be spuriously found in data whose underlying distribution does not have those features.
To do this, we construct a null distribution by simulating BBH observations that would occur if the underlying astrophysical distribution was a single power law with no substructure in a finite range.
The procedure for creating synthetic BBH observations is described in Appendix~\ref{appendix:mockPE}.
Mock observations are combined with corresponding sensitivity estimates in a hierarchical Bayesian analysis, described in \cite{loredo_handling_2009,mandel_extracting_2019,thrane_introduction_2019}.
We analyze these simulations in the same way as the BBHs in GWTC-3 to determine how often the features observed in GWTC-3 would be found from an underlying distribution without those features.

\subsection{\PS{} Mass Model}
\label{sec:methods-PS}
We use the \PS{} semi-parametric primary mass model as a flexible model that is easily capable of finding peaks and valleys in the mass distribution \citep{Edelman_2022, O3b_pop}.
This model parameterizes perturbations or deviations from a simpler underlying distribution with flexible cubic spline functions. Specifically, given an underlying hyper-prior for primary mass, $p(m_1 | \Lambda)$, the \PS{} model describes the primary mass distribution as:
\begin{align}
\begin{aligned}
    p_\mathrm{spline}&(m_1 | \Lambda, \{m_i\}, \{f_i\}) \\
    &\propto p(m_1 | \Lambda) \exp(f(m_1 | \{m_i\}, \{f_i\}))
    \end{aligned}
\end{align}
where $f(m_1| \{m_i\}, \{f_i\})$ is the function describing the perturbations, which we model with a cubic spline function interpolated by introduced hyper-parameters, $\{m_i\}$, the locations of spline knots in mass space, and $\{f_i\}$, the height of the perturbation function at each knot.
This describes a semi-parametric model as it includes a simple ``parametric" component (the underlying distribution) in addition to a non-parametric component that models the perturbation around the simple description.
For this study we use the simplest primary mass model for the underlying description, which is the \textsc{Truncated} model, describing a power law with sharp cutoffs at the lower and upper mass bounds \citep{2017ApJ...851L..25F, Edelman_2022}.
While this model has been shown to insufficiently describe the primary mass distribution, it captures the majority of the broadest features \citep{O3a_pop, O3b_pop}.

To assess the significance of peaks or valleys found with the \PS{} model one can look at the posterior distribution of the perturbation heights as a function of mass.
This tells us how far ``off'' the simple power law description is from accounting for the data.
Specifically we can find what percentile $f(m_1) = 0$ falls in the posterior distribution as a function of mass.
For data exactly distributed as a power law (the underlying population), the inferred perturbation function should be symmetric about 0 with widths determined by the prior distributions on the knot heights \new{and the number of observed events}.
At masses where the percentile of zero perturbation approaches 100\% (0\%) we can say there is an over (under) density of events at these masses, compared to the underlying power law distribution.
This is identical to the analysis done by \cite{O3b_pop}, who use the percentile at which the perturbation function excludes zero at a given location as a metric for how significant a feature is at that location.

\subsection{Metrics of Feature Significance}
\label{sec:methods-metrics}

As described in Section~\ref{sec:methods-PS}, the \PS{} model makes use of a perturbation function constructed from cubic splines.
The height of the perturbation function, $f(m_1)$, at a point in primary mass, $m_1$, is then a direct measure of the deviation from a power law at that point.
We can determine how often one would find spurious evidence for substructure by simulating catalogs from a power law, fitting them with the \PS{} model, and examining the resulting perturbation function.

If the mock catalogs produce perturbation functions with similar amplitudes to those seen for GWTC-3, the structure in the GWTC-3 fit might be described by Poisson noise.
On the other hand, if the perturbation functions produced by fits to the mock catalogs are always lower in amplitude to that of the GWTC-3 fit, the structure in the GWTC-3 data is likely to be present in the underlying distribution. 

For a given mock catalog, we find the $m_1$ value where the \new{median \emph{a posteriori} value of the} perturbation function is maximal.
We obtain the posterior distribution of perturbation function amplitudes at that location, $g(f_{\max})$.
We repeat this for all mock catalogs, obtaining a set of maximal perturbation function distributions, $\{g_j(f_{\max})\}$.
These are plotted in light grey on the left panels of Figure~\ref{fig:three-largest-perturbations}.
The locations of the three maximal perturbation function amplitudes in the GWTC-3 fit are, from least to most Bayesian significance, \result{$13.8\Msun, 10.3\Msun$, and $35.7\Msun$}.
The posterior distributions of perturbation function heights at these locations are $g_{\text{GWTC-3}}(f(13.8\Msun)), g_{\text{GWTC-3}}(f(10.3\Msun))$, and $g_{\text{GWTC-3}}(f(35.7\Msun))$, and are plotted in orange in the left panels of Figure~\ref{fig:three-largest-perturbations}.
The amplitude of the perturbation function at \result{$13.5\Msun$} is negative (i.e. it is a dip rather than a bump), so we flip its distribution about zero for more direct comparison.
The same is done for all $g(f_{\max})$ whose medians are negative, as the perturbation function's prior is symmetric about zero.

\subsection{Simulation Study}
\label{sec:results-2}
\begin{figure*}
    \centering
    \includegraphics[width=\textwidth]{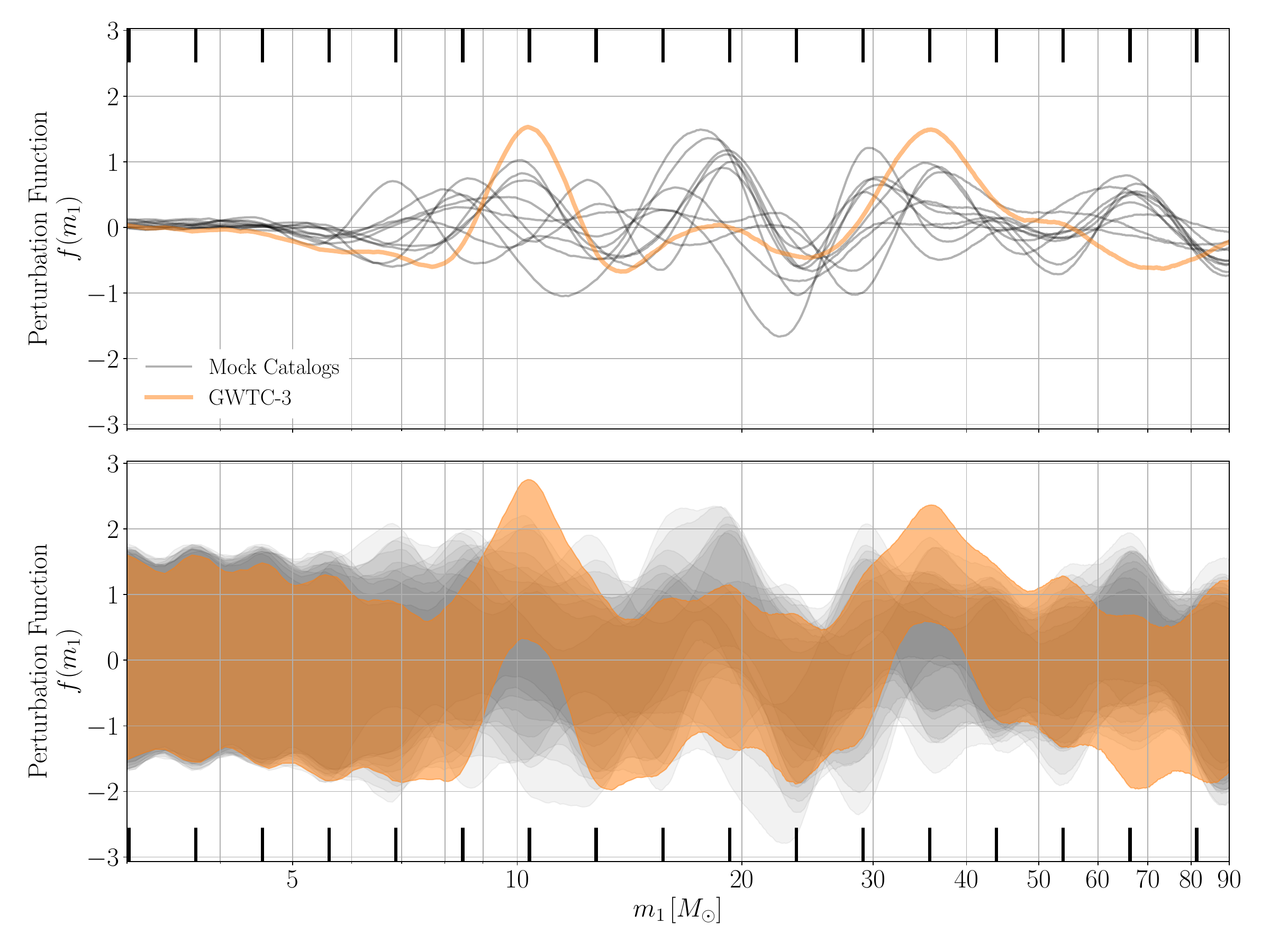}
    \caption{Median (\emph{top panel}) and 90\% credible interval (\emph{bottom panel}) of the perturbation function resulting from the \PS{} fit to the primary masses in GWTC-3 (\emph{orange}) and in \result{10} mock catalogs (\emph{grey}).
    The perturbation function multiplies a smoothed power law in primary mass to add modulations to an otherwise monotonic distribution, making it a direct measure of deviations from a power law.
    It is a cubic spline with knots fixed at the locations indicated by the black vertical tick marks.
    The prior on the perturbation heights is the unit normal distribution, as can be seen below $\sim5\Msun$ where there are no detections to constrain the likelihood and the posterior reverts to the prior.
    The perturbation function corresponding to GWTC-3 events appears large in amplitude in three locations: $\sim 10 \Msun$, $\sim 14 \Msun$, and $\sim 35 \Msun$.
    While the medians of the perturbation function at these distributions are comparable in amplitude, the posterior distribution at $\sim 35 \Msun$ ($\sim 14 \Msun$) is the most (least) tightly constrained.
    }
    \label{fig:spline_example}
\end{figure*}

To determine whether the features in the mass spectrum of GWTC-3 BBHs are the result of Poisson noise of a finite number of observations drawn from a featureless power law, we compare \PS{} fits using the GWTC-3 catalog and \result{300} mock catalogs generated from a ``featureless'' power law.
The mock catalogs considered in this section are all generated from the same underlying distribution: a truncated power law in primary mass, mass ratio, and redshift, with a smoothing at low component masses to ensure the peak of the mass distribution is not in the same location as the minimum mass.
The explicit form of the mock catalogs' population model, including values of all of its hyperparameters, can be found in Appendix~\ref{appendix:MDC}.

\begin{figure*}
    \centering
    \includegraphics[width=\textwidth]{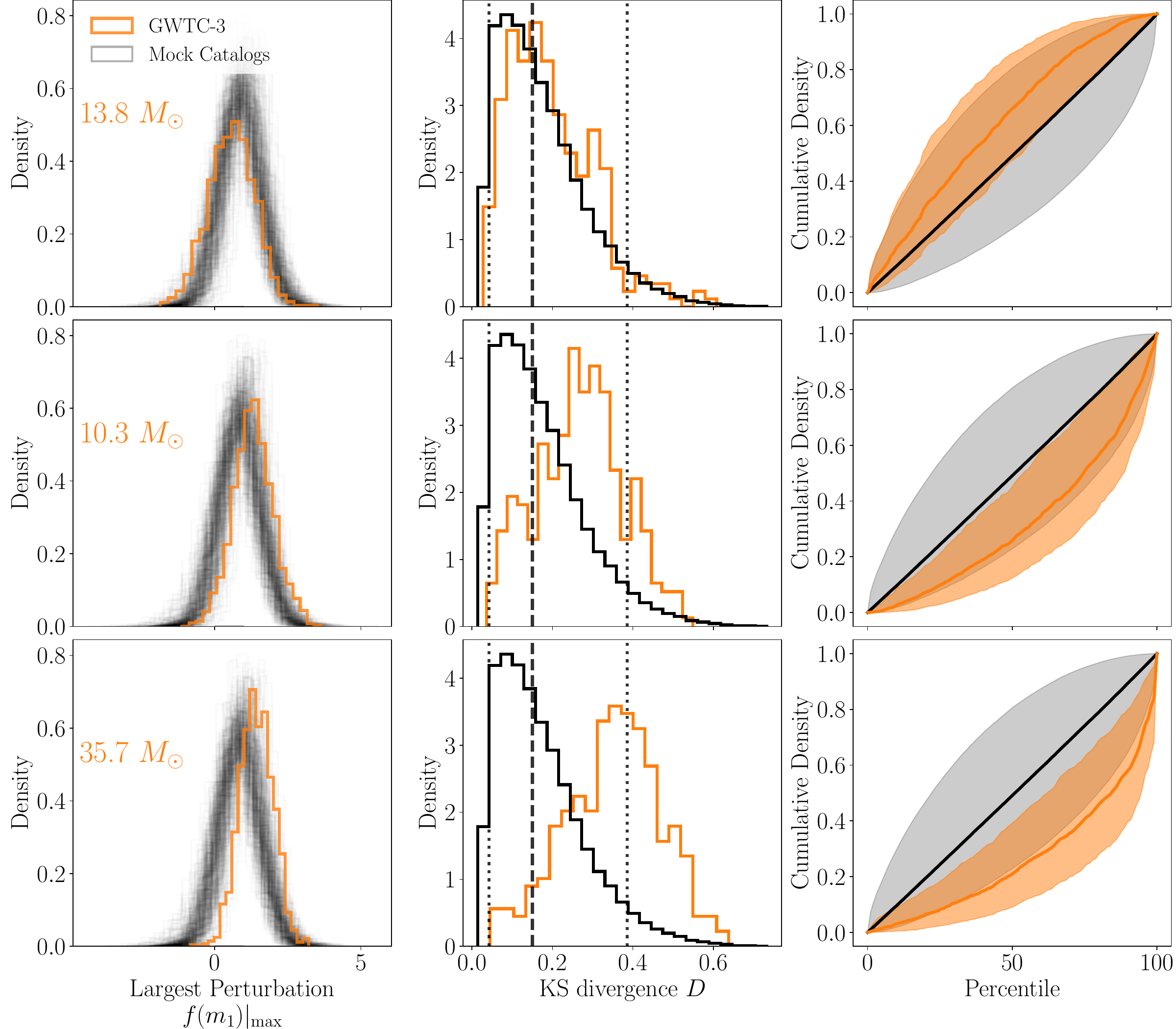}
    \caption{Three largest deviations from a power law observed in GWTC-3 compared to mock catalogs.
    \emph{Left column:} The posterior distribution of perturbation function heights at the location where the posterior distribution is maximal for mock catalogs (\emph{light grey}) and GWTC-3 (\emph{solid orange}).
    \emph{Middle column:} Null distribution (\emph{black}) and GWTC-3 distribution (\emph{orange}) of Kolmogorov–Smirnov (KS) divergences between the individual distributions in the left column.
    Smaller values of the KS divergence indicate more similar distributions. 
    \emph{Right column:} Null distribution (\emph{black}) and GWTC-3 distribution (\emph{orange}) of percentiles.
    Large deviations from the diagonal indicate a more significant rightward shift of the GWTC-3 distribution relative to the mock catalogs.
    Each row corresponds to a different local extremum for GWTC-3: \result{$m_1= 13.8\Msun$} (\emph{top}), \result{$m_1= 10.3\Msun$} (\emph{middle}), and \result{$m_1= 35.7\Msun$} (\emph{bottom}), while the global extrema for each mock catalog are shown in all rows, along with the aggregated distribution across all mock catalogs (\emph{solid black}).
    The $\sim35\Msun$ peak is an outlier with respect to both the KS and percentile statistics, but the other two features are more ambiguous.
    }
    \label{fig:three-largest-perturbations}
\end{figure*}

\new{Full parameter estimation was not performed on each event in each mock catalog; instead, we use prescriptions for generating event posteriors that reproduce the correlations between an event's parameters, as well as the typical uncertainties seen in GWTC-3.
We show in Appendix~\ref{appendix:mockPE_accuracy} that the prescriptions used are sufficient to reproduce population analyses such as the ones scrutinized in this work.}
\new{In order for our comparisons to be consistent between featureless mock catalogs and GWTC-3, we recreate GWTC-3's event posteriors with the same prescriptions as were used for the mock catalogs, perform a population analysis on those, and use the resulting perturbation function for all comparisons to mock catalogs. \new{Our population re-analysis of the GWTC-3 events with mock posteriors appears consistent with the full analysis presented by the LVK in \citealt{O3b_pop}.}
Lines labeled ``GWTC-3'' or depicted in orange in Figures ~\ref{fig:spline_example},~\ref{fig:three-largest-perturbations}, and \ref{fig:zero_exclusion} refer to the analysis done on the recreated version of GWTC-3.
All analyses presented here were repeated using the original LVK-released parameter estimation samples in Appendix~\ref{appendix:mockPE_accuracy}, and the same qualitative conclusions were reached, though with slightly more statistical significance.}

Despite knowing the parameters of the underlying population for the mock catalogs, we allow all hyperparameters to vary when fitting \PS{} to the mock catalogs.
The resulting perturbation functions are shown in Figure~\ref{fig:spline_example} for \result{10} randomly chosen mock catalogs and GWTC-3.
The perturbation functions deviate from their prior distribution in the mass range where detections exist (above $\sim5\Msun$ and below $\sim85\Msun$), even in the case of mock catalogs.
This means that the perturbation functions are informed by the mock data despite the mock data not inherently requiring a deviation from a power law.
The question still remains whether the perturbation function heights inferred from mock catalogs with no substructure are larger than those inferred from GWTC-3.
While nonzero values of the perturbation function are common in the \result{10} mock catalog fits shown in Figure~\ref{fig:spline_example}, only a few amplitudes appear comparable in height to the three largest amplitudes of the GWTC-3 perturbation function.

To verify this, we isolate the largest amplitude perturbations for all \result{300} mock catalog fits and compare them to the three largest amplitude perturbations for the GWTC-3 fit.
These are plotted in the leftmost panels of Figure~\ref{fig:three-largest-perturbations}.
The light grey curves are the posterior distributions of largest perturbation function amplitudes $\{g_j(f_{\max})\}$ for each simulated catalog $j$.
These appear to have the same general shape as one another, though with noticeable scatter.
The orange curves in each panel are the posterior distributions of GWTC-3's perturbation function $g_{\text{GWTC-3}}(f(m_1))$ at its three maximal locations: \result{$m_1= 13.8\Msun, 10.3\Msun,$ and $ 35.7\Msun$}.

The distribution for the $\sim 14\Msun$ dip appears qualitatively similar to that of the simulated catalogs, the $\sim 10\Msun$ peak appears to be slightly shifted with respect to most of the simulated catalogs but still within their range, and the $\sim 35\Msun$ peak is noticeably shifted towards higher values relative to the bulk of the simulated catalog distributions.
This suggests that the $\sim 35\Msun$ peak is unlikely to be the result of Poisson noise or modeling systematics, while other features could plausibly be explained by those effects. 

\subsubsection{Maximum Perturbation Amplitude}
\label{sec:ks_perc}

To obtain a more quantitative measure, we derive several metrics from the distributions of maximal perturbation function amplitudes.
The first uses the Kolmogorov–Smirnov (KS) test: we compute the KS divergence $D$ between each of the $\{g_j(f_{\max})\}$ distributions 
to obtain a null distribution of KS divergences, shown in the solid black curve in the middle column of Figure~\ref{fig:three-largest-perturbations}.
We then perform a KS test between the $\{g_j(f_{\max})\}$ distributions and $g_{\text{GWTC-3}}(f(m_1))$ 
and obtain the orange curves in the middle column of Figure~\ref{fig:three-largest-perturbations}.
From this, we find that the KS divergences for mock catalogs are larger than those of GWTC-3 \result{$20\%$, $11\%$, and  $7\%$} of the time for the $14\Msun$, $10\Msun$, and $35\Msun$ features, respectively.
This means, for example, that mock catalogs can produce perturbation function posteriors as tall as the one inferred from GWTC-3 at $\sim35\Msun$ only $7\%$ of the time.
Written in terms of $g(f)$, \result{$g_{\text{GWTC-3}}(f(14\Msun)) \neq g_j(f_{\max})$ to $20\%$, $g_{\text{GWTC-3}}(f(10\Msun)) \neq g_j(f_{\max})$ to $11\%$, and $g_{\text{GWTC-3}}(f(35\Msun)) \neq g_j(f_{\max})$ to $7\%$}.
Though none of these percentages are convincingly small, this indicates that the orange histograms are more statistically distinct from the black histograms in the case of the $\sim35\Msun$ peak than they are in the cases of the features at $10\Msun$ and $14\Msun$.

The second metric is obtained by quantifying the shift of $g_{\text{GWTC-3}}(f(m_1))$ relative to the set of $\{g_j(f_{\max})\}$.
For each point in $g_{\text{GWTC-3}}(f(m_1))$, we calculate the percentile in which it lies in each of the $\{g_j(f_{\max})\}$, obtaining the orange bands in the rightmost panels of Figure~\ref{fig:three-largest-perturbations}.
For comparison, we do the same for each of the $\{g_j(f_{\max})\}$ relative to each other, constructing the grey bands in the rightmost panels of Figure~\ref{fig:three-largest-perturbations}.
We then take the mean of the set of light orange bands and light black bands to obtain the solid orange and solid black curves, respectively.
The black bands serve as null distributions, so large deviations from those indicate significant shifts.
We observe a large deviation for the $\sim35\Msun$ peak, a moderate deviation for the $\sim10\Msun$ peak, and only a slight deviation for the $\sim14\Msun$ dip.
Quantitatively, \result{$g_{\text{GWTC-3}}(f(35\Msun)) \geq g_j(f_{\max})$ to $83^{+17}_{-69}\%$} (90\% credible interval), meaning that the $\sim35\Msun$ peak lies in the \result{$83^{+17}_{-69}$}rd percentile of the mock catalogs' largest perturbation heights.
For the other features, \result{$g_{\text{GWTC-3}}(f(10\Msun)) \geq g_j(f_{\max})$ to $74^{+25}_{-60}\%$ and $g_{\text{GWTC-3}}(f(14\Msun)) \geq g_j(f_{\max})$ to $34^{+52}_{-32}\%$.}
In comparison, the corresponding statistic for the null distributions is $g_j(f_{\max}) \geq g_i(f_{\max})$ to $50^{+47}_{-46}\%$.

It is not possible to draw firm conclusions from these large uncertainties. 
However, the central values indicate that the $\sim35\Msun$ peak is noticeably shifted relative to the mock catalogs' perturbation functions, \new{the $\sim10\Msun$ peak is moderately shifted, and the $\sim14\Msun$ dip even has a slightly lower amplitude than the maximum perturbation functions typical of mock catalogs}.

\subsubsection{Inconsistency With a Power Law}
\label{sec:zero-exclusion}
\begin{figure}
    \centering
    \includegraphics[width=\columnwidth]{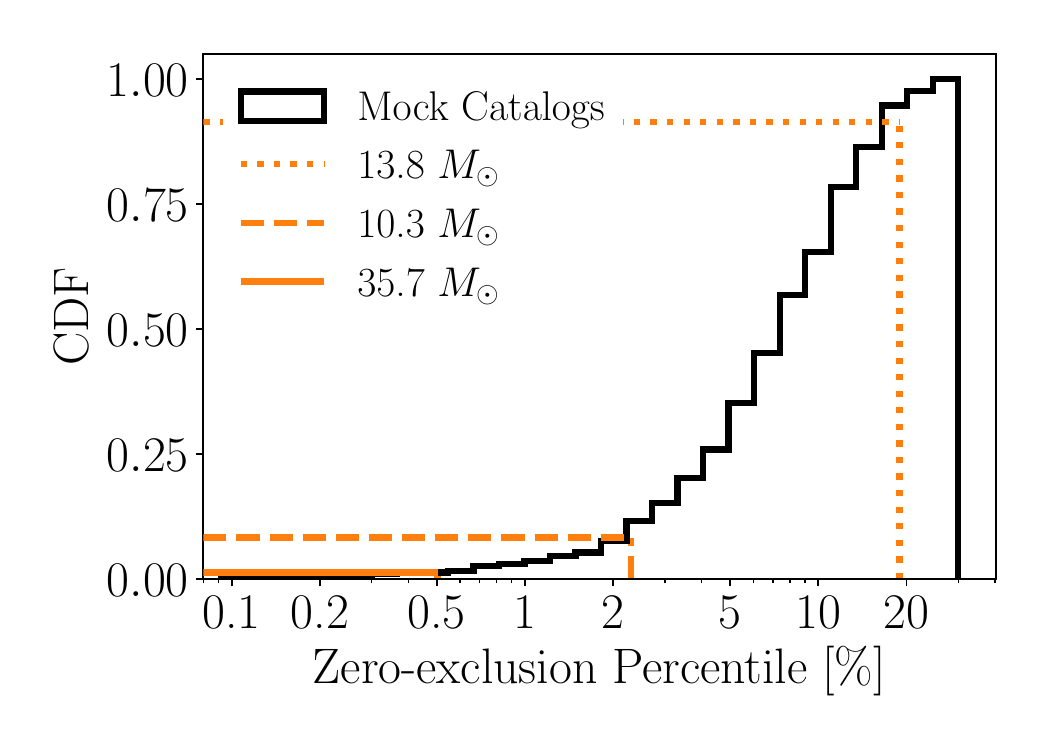}
    \caption{Percentile at which the posterior distribution of the perturbation function excludes zero for GWTC-3 (\emph{orange vertical lines}) and catalogs drawn from a featureless distribution (\emph{black histogram}).
    For GWTC-3, we evaluate the perturbation function's posterior distribution at primary mass ($m_1$) values of  $13.8\Msun$ (\emph{dotted}), $10.3\Msun$ (\emph{dashed}), and $35.7\Msun$ (\emph{solid}).
    For mock catalogs, we find the primary mass value at which the perturbation function is maximal and evaluate its posterior distribution there.
    The values reported here are the percentage of the posterior distribution that is greater than zero at those values in $m_1$.
    The $13.5\Msun$ feature excludes zero to a level comparable to some of the mock catalogs, but the other two features exclude zero to a level not reproducible by any mock catalogs.
    }
    \label{fig:zero_exclusion}
\end{figure}

The final metric we consider is inspired by the statistic presented in \cite{O3b_pop}, which states that ``the inferred perturbation $f(m_1)$ strongly disfavors zero at both the $10\Msun$ and $35\Msun$ peak.''
We therefore turn from considering the full distribution of perturbation function heights at a given location to the percentile at which it excludes zero.
A perturbation function amplitude of zero is a useful reference point for several reasons.
The most intuitive is that it causes the population model to behave like a featureless power law, so a posterior that excludes zero to high credibility indicates an inconsistency with a power law.
Zero is also the mean of the prior predictive distribution for the perturbation function: the prior allows for equal upwards and downwards fluctuations, symmetric about zero perturbation.
Similarly, a vanishing perturbation function amplitude is the state to which we expect the posterior predictive distribution to asymptote in the limit of infinite detections from an underlying power law distribution.
We therefore plot the percentile at which each mock catalog excludes zero perturbation in Figure~\ref{fig:zero_exclusion}.

We then calculate how often a simulated catalog's perturbation function excludes zero to the same credibility as that of GWTC-3.
\new{This is the same as finding the point along the $y$-axis of Figure~\ref{fig:zero_exclusion} at which each of the vertical orange lines hits the cumulative distribution function.}
\new{\result{1.7\%, 10.0\%, and 92.7\%} of the $\{g_j(f_{\max})\}$ exclude zero to the same percentile as $g_{\text{GWTC-3}}(f(35\Msun))$, $g_{\text{GWTC-3}}(f(10\Msun))$, and $g_{\text{GWTC-3}}(f(14\Msun))$, respectively}.
The fact that, \new{for example,} $g_{\text{GWTC-3}}(f(14\Msun)) < 0$ to $20.7\%$ but $92.7\%$ of mock catalogs have a similar or smaller \new{statistical excursion} is due in part to the difference between Bayesian credible intervals and frequentist $p$-values, and because our metric corrects for the look-elsewhere effect by comparing GWTC-3's perturbation function at specific locations to all possible locations in the mock catalogs.

Combined with the metrics presented in Section~\ref{sec:ks_perc}, the results above lead us to conclude that the peak at ${\sim35\Msun}$ is difficult to reproduce with featureless catalogs, but it is possible that the dip at $\sim14\Msun$ is just a large fluctuation rather than a astrophysical feature.
\new{The peak at ${\sim10\Msun}$ is difficult to reproduce with featureless catalogs, though it is easier to reproduce than the ${\sim35\Msun}$ peak.
We discuss the interpretation of this feature in more detail in Section~\ref{sec:discussion}.}

In summary, featureless catalogs can sometimes produce features as tall as the $\sim10\Msun$ peak, and they can sometimes produce perturbations constrained away from zero with the same \new{credibility}.
The dip at $\sim14\Msun$ could be a Poisson fluctuation because fits to featureless catalogs can easily produce perturbations as large, and as \new{credibly} constrained away from zero perturbation.
The peak at $\sim35\Msun$ is difficult to reproduce by mock catalogs in any way: its perturbation amplitude is too large and too \new{credibly} constrained away from zero.

The fact that we find one of the features explainable by Poisson noise is consistent with Section~\ref{sec:results-1}, which suggests that up to two of the excursions from a power law can be explained by Poisson fluctuations.
Our conclusions are also in broad agreement with those presented in \cite{O3b_pop}, as they report confident detections for the two largest peaks in the mass distribution but only modest evidence for the dip at $\sim14\Msun$.

\comment{If the features at $\sim10\Msun$, $\sim14\Msun$, and $\sim35\Msun$ are in fact true features of the underlying distribution, by the end of the LVK's fourth observing run the perturbation function will be constrained away from zero at \result{X\%, Y\%, and Z\%}, respectively. 
We will therefore be able to rule out the possibility of Poisson noise for all three features, including the dip at $\sim14\Msun$.}

\section{Discussion}
\label{sec:discussion}
Previous analyses of the BBH mass spectrum by the LVK and others have found evidence for structure beyond a simple power law \citep{O3a_pop, O3b_pop}.
There has been considerable work exploring possible astrophysical causes of these identified features.
Our aim is instead to determine, from a statistical viewpoint, whether astrophysical arguments need be invoked at all.

We first demonstrate that it is only possible for up to two of the three deviations from a power law to be explained by Poisson noise about a single power law distribution.
Therefore, at least one feature must be added on top of a power law to describe the data.

We then perform a more thorough analysis, simulating thousands of BBHs with measurement uncertainty, selection effects, and a known underlying distribution.
We fit the \PS{} model to the resulting catalogs and find that the data is inconsistent with a single power law, agreeing with the LVK result. 
However, we find that one of the previously identified features, an underdensity at $\sim14\Msun$, may not be present in the true astrophysical distribution.
Instead, it may have been the result of a Poisson fluctuation around a simple power law, or an artifact of the models used to fit the mass spectrum.
The metrics constructed in this work differ from those previously used to assess the significance of features in the mass distribution because, by virtue of comparing to several simulated catalogs, they correct for the look-elsewhere effect.
This is only in mild tension with the conclusions reached by \cite{O3b_pop}, as they report ``modest evidence'' in favor of a dip at $14\Msun$.

We find the other two previously identified peaks, at ${\sim10\Msun}$ and ${\sim35\Msun}$, unlikely to be the result of Poisson noise or modeling artifacts.
Simulated catalogs coming from distributions that do not include these features can reproduce the height of the ${\sim10\Msun}$ peak, but not its lack of support for zero perturbation.
The ${\sim35\Msun}$ peak is difficult to reproduce from featureless catalogs in any way.

Our conclusions are consistent with a recent study by \cite{callister_parameter-free_2023} who fit the BBH mass distribution with an autoregressive model and find that the primary mass distribution gradually decreases as a function of mass and exhibits two local maxima \new{with a relatively flat continuum between them. They interpret the 14 solar mass dip found by other analyses to be  a flattening of the power law index at lower masses rather than a local minimum.}
We also find similar results to \cite{edelman_cover_2022} who construct the mass distribution entirely from basis splines and find peaks at $\sim10\Msun$ and $\sim35\Msun$. 
The significance of the peaks near $10\Msun$ and $35\Msun$, as well as the lack of significance of the dip near $14\Msun$, is also in agreement with \cite{Sadiq_2022} and \cite{wong_automated_2022}.

The dip near $\sim14\Msun$ may be a large Poisson fluctuation or an artifact of the models used to characterize it.
If it is in fact a feature of the underlying distribution, it is difficult to resolve with current observations.
\comment{However, it will be distinguishable from noise by the end of the fourth LVK observing run to \result{X\%} if it is present in the true distribution.}

The peak near $\sim10\Msun$ is likely an imprint of the true astrophysical distribution.
\new{Its amplitude is slightly larger than what featureless catalogs can produce with random fluctuations, and it is inconsistent with the power law that describes the rest of the BBH mass distribution at a level that only a small fraction of featureless catalogs can achieve.
We therefore report moderate evidence that} additional structure beyond a power law is needed to explain the peak at $\sim10\Msun$.

The $\sim10\Msun$ feature may either be an additional peak that is distinct from the one created by the underlying smoothed power law at $\sim7\Msun$ \citep{O3b_pop, Edelman_2022, tiwari_exploring_2022} or the sole peak in the region between $\sim5\Msun$ and $\sim20\Msun$ \citep{edelman_cover_2022}.
These two possibilities can be seen in Figure~\ref{fig:dR_dm1}. 
The former scenario is the case where we interpret the first two peaks in the orange band as distinct from one another, therefore treating the global maximum inferred by \PS{} as a different feature from the global maximum inferred by \textsc{Power Law + Peak}.
In the latter scenario, the role of the perturbation function is to shift the global maximum from the value inferred by the power law component to a slightly higher value without removing the mass distribution's support for $5$--$10\Msun$ objects.
A simple smoothed power law, such as that employed by the \textsc{Power Law + Peak} model (see grey and blue bands in Figure~\ref{fig:dR_dm1}), may not be flexible enough to place a global maximum at $\sim10\Msun$ while also fitting the correct slope at larger masses and fitting the correct merger rate below $\sim10\Msun$, so it places its global maximum at $\sim7\Msun$.
This scenario, in which there is a single local maximum below $\sim12\Msun$, is consistent with \cite{edelman_cover_2022} and \cite{callister_parameter-free_2023}, both of whom find only one significant maximum between approximately \result{$3\Msun$ and $12\Msun$} using fully non-parametric methods.
If this interpretation is correct and the global maximum of the BBH mass distribution is indeed offset from the minimum mass by $\sim5\Msun$, the upper edge of the lower mass gap may not be as morphologically simple as previously assumed \citep[e.g.,][]{fishbach_does_2020, farah_bridging_2022,ezquiaga_spectral_2022}, making it potentially difficult to resolve with parametric models alone.
\new{The marginal evidence for the significance of the $\sim10\Msun$ peak is likely driven by the fact that current observations are insufficient to distinguish between these two scenarios, which is unsurprising considering the lower sensitivity of GW detectors to low-mass events relative to high-mass events.}

\new{A peak anywhere between $\sim 7\Msun$ and $\sim 10\Msun$} could be indicative of particular evolutionary processes that are dominant within formation environments. 
\cite{van_son_no_2022} showed that a global maximum near this value is consistent \new{with} and robustly predicted by the stable mass transfer channel in isolated binary evolution, as stability during mass transfer requires mass ratios between the donor star and accreting compact object to be relatively symmetric, and stellar companions to $\sim 10~\Msun$ BHs must be near this mass to form compact objects above the minimum BH mass. 
This may be an indication that the stable mass transfer channel operates more efficiently than the traditional common envelope channel for generating merging BBHs. 
\new{If the stable mass transfer channel is indeed the cause of the global maximum in the primary mass distribution, the exact location of this global maximum will constrain the core mass fraction, mass transfer stability, and mass transfer efficiency of this process \citep{van_son_no_2022}}.
Though dynamical formation channels with low escape velocities, such as globular clusters, struggle to produce a global maximum at $10~\Msun$~\citep{antonini_coalescing_2022}, dynamical environments with higher escape velocities may more readily produce merging BBHs with lower masses around $10~\Msun$ due to the more prevalent lower-mass BHs preferentially remaining bound to these clusters following supernova kicks.

We find that the peak centered on $35\Msun$ is the most likely to be a feature of the true underlying distribution.
This bodes well for the ``spectral siren'' \citep{farr_future_2019, ezquiaga_spectral_2022} method of estimating cosmological parameters from GW observations, as this peak happens to be the most informative feature for this method since it is a well-measured, somewhat sharp feature in the mass distribution~\citep{gwtc3_cosmo}.
The astrophysical process that gives rise to this feature is still a topic of discussion.
The key reason for including a flexible bump-like feature in the phenomenology of parametric models, such as the \textsc{Power law + Peak} model used by the LVK \citep{talbot_measuring_2018}, was to accommodate a potential build-up of BHs with masses just below the pair instability mass gap, as pulsational pair instability supernovae are predicted to efficiently shed material from high-mass stars with cores in the mass range of $M_\mathrm{core} \sim 45 - 65$~\citep{woosley_pulsational_2017,woosley_evolution_2019,marchant_pulsational_2019,renzo_predictions_2020}. 
It is difficult to reconcile the locations of the local maxima found in the BBH primary mass distribution with predictions of the pair instability process in the cores of massive stars. 
The largest uncertainty determining the location of the lower edge of the pair instability mass gap is the $^{12}\mathrm{C}(\alpha,\gamma)^{16}\mathrm{O}$ reaction rate, which determines the abundance of oxygen in stellar cores~\citep[e.g.,][]{farmer_mind_2019}. 
Higher $^{12}\mathrm{C}(\alpha,\gamma)^{16}\mathrm{O}$ reaction rates lead to a higher oxygen abundance in the stellar core, which will ignite explosively during core collapse and lead to (pulsational) pair instability supernovae occurring at lower core masses. 
However, even at $3 \sigma$ deviations above the median measured value of the $^{12}\mathrm{C}(\alpha,\gamma)^{16}\mathrm{O}$ reaction rate, the lower end of the mass gap only reaches $\approx 38~\Msun$~\citep{farmer_constraints_2020}.
This is above where the measured overdensity in the observed mass spectrum occurs. 
This may be an indication that the peak at $35~\Msun$ is the result of \new{certain isolated binary evolution scenarios \citep[e.g. chemically homogeneous evolution, see ][]{du_buisson_cosmic_2020,zevin_one_2021,bavera_probing_2022}}, another BBH formation channel\new{ entirely} (e.g. globular clusters, see \citealt{antonini_coalescing_2022}), or that stellar evolution models are missing particular ingredients that can shift the location of the pair instability gap (relaxing the assumption that the exploding stars are hydrogen-free, adjustments to convective overshooting, see e.g. \citealt{iorio_compact_2022}).

Additionally, several studies have suggested that the observed peaks in the BBH mass distribution can be explained by successive generations of hierarchical mergers \citep{Tiwari_2021,mahapatra_black_2022,tiwari_exploring_2022}, though no correlation has been detected in the spin distribution of BBHs~\citep{biscoveanu_binary_2022}, which is also necessitated by the hierarchical merger formation scenario \citep{gerosa_are_2017, fishbach_are_2017, rodriguez_black_2019, kimball_black_2020,doctor_black_2020,doctor_black_2021,gerosa_high_2021}.
Additionally, for these peaks to correspond to hierarchical mergers of the same population, the dominant hierarchical pairing would have to be the first generation BH with a third generation BH \citep{mahapatra_black_2022, tiwari_exploring_2022}, whereas the dominant pairing predicted by \cite{rodriguez_black_2019} is a first generation BH generation with a second generation BH.
While it is certainly possible that GWTC-3 contains hierarchical mergers (e.g. \citealt{abbott_gw190521_2020}, though also see \citealt{fishbach_minding_2020}), the relative fraction of events formed this way is likely too small to form the structure observed in the primary mass distribution \citep{kimball_evidence_2021}, and some fine-tuning may be needed to avoid a cluster catastrophe~\citep{2022ApJ...935L..20Z}.
The exact physical reason for the overdensity at $35~\Msun$ therefore remains unclear.
However, we confirm that it is a robust signature in the observational data; future observing runs will help to constrain its precise location, width, and possible redshift evolution.


\section{Acknowledgements}
The authors gratefully acknowledge Reed Essick for helpful insights on injections and model systematics, as well as Thomas Callister and Thomas Dent for useful comments on the manuscript.
\new{We also thank the anonymous reviewer for suggesting several helpful checks of the analysis presented here.}
A.M.F. is supported by the National Science Foundation Graduate Research Fellowship Program under Grant No. DGE-1746045.
B.E and B.F are supported in part by the National Science Foundation under Grant PHY-2146528. 
M.Z. is supported by NASA through the NASA Hubble Fellowship grant HST-HF2-51474.001-A awarded by the Space Telescope Science Institute, which is operated by the Association of Universities for Research in Astronomy, Inc., for NASA, under contract NAS5-26555.
J.M.E. is supported by the European Union’s Horizon 2020 research and innovation program under the Marie Sklodowska-Curie grant agreement No. 847523 INTERACTIONS, by VILLUM FONDEN (grant no. 37766), by the Danish Research Foundation, and under the European
Union’s H2020 ERC Advanced Grant “Black holes: gravitational engines of discovery” grant agreement no. Gravitas–101052587.  
D.E.H is supported by NSF grants AST-2006645 and PHY-2110507, as well as by the Kavli Institute for Cosmological Physics through an endowment from the Kavli Foundation and its founder Fred Kavli.
D.E.H also gratefully acknowledges the Marion and Stuart Rice Award.
This material is based upon work supported by NSF LIGO Laboratory which is a major facility fully funded by the National Science Foundation. This research has made use of data, software and/or web tools obtained from the Gravitational Wave Open Science Center 
(\url{https://www.gw-openscience.org/}), a service of LIGO Laboratory, the LIGO Scientific Collaboration and the Virgo Collaboration. 
The authors are grateful for computational resources provided by the LIGO Laboratory and supported by National Science Foundation Grants PHY-0757058 and PHY-0823459.  
This work benefited from access to the University of Oregon high performance computer, Talapas.

\software{\texttt{gwpopulation}~\citep{talbot_parallelized_2019}, 
    \texttt{bilby}~\citep{Ashton_2019, bilby_gwtc1},
    \texttt{dynesty}~\citep{speagle_dynesty_2020},
    \texttt{numpy}~\citep{harris_array_2020}, 
    \texttt{xarray}~\citep{hoyer_xarray_2017}, 
    \texttt{matplotlib}~\citep{hunter_matplotlib_2007}, 
    \texttt{pandas}~\citep{reback2020pandas}
}

\bibliography{references}{}
\bibliographystyle{aasjournal}

\appendix

\section{Generation of Mock Observations in \texttt{GWMockCat}}
\label{appendix:mockPE}
We describe the process used to simulate gravitational wave event posteriors in mass and redshift, based on the procedure developed in \cite{FFH}.

This process neglects the generation of spin posteriors as this work only seeks to understand the significance of features in the mass distribution, and individual-event likelihoods are approximately separable in spin and primary mass for BBHs, and we do not model any spin populations in this work.
However, spin and mass parameters are not totally uncorrelated for low-mass or high mass ratio events, so future work attempting to validate features seen in the mass ratio distribution, NSBH, or BNS populations should consider simulating spin parameters as well.
A lightweight, publicly-available \texttt{python} package that can reproduce these mock posteriors and generate similar catalogs from arbitrary underlying populations and detector sensitivities is available for download and installation\footnote[4]{ \url{https://git.ligo.org/amanda.farah/mock-PE}}.
The package is called \texttt{GWMockCat}, and installation instructions, examples, and documentation are available in the git repository.
Several packages exist to draw events from BBH population models \citep[][e.g.]{belczynski_compact_2008,breivik_cosmic_2020,riley_rapid_2022}, some of which also simulate GW detector selection effects \citep{karathanasis_gwsim_2022}.
\texttt{GWMockCat} complements these by additionally simulating event-level posteriors without the need to run full parameter estimation inference, saving significant computational time.

To create realizations of catalogs that would \new{reasonably} result from a known underlying astrophysical population, $p(m_1,m_2,z)$, we first make independent draws of the event parameters, $\{m_1, m_2, z\}$, from that population model.
Each draw corresponds to a potential event in the catalog, although we draw many more potential events than we wish to keep since not all events generated from the astrophysical distribution will ultimately be detected.
We then convert each event's redshift $z$ and source-frame component masses to a detector-frame (redshifted) chirp mass, $\mathcal{M}_{\text{det}}$, and symmetric mass ratio, $\eta$. 
The symmetric mass ratio and source-frame chirp mass $\mathcal{M}$ are related to the source-frame component masses via
\begin{align}
    \eta &= \frac{m_1 m_2}{(m_1 + m_2)^2} \\
    \mathcal{M} &= \frac{(m_1 m_2)^{3/5}}{(m_1 + m_2)^{1/5}} .
    \label{eq:mass-transforms}
\end{align}
All detector-frame masses are related to their source-frame values via $M_{\det} = M (1+z)$, where $M$ can describe any parameter with units of mass (e.g. $m_1, m_2,$ or $\mathcal{M}$).

We then utilize the basic procedure outlined in \cite{2018ApJ...863L..41F} and \cite{FFH} to assign ``observed" parameters for each event, using measurement uncertainty \new{that is correlated across parameters} and a mock parameter estimation likelihood. 
We first calculate an optimally oriented signal-to-noise ratio (SNR) $\rho_{\text{opt}}$ from the true event parameters using a characteristic power spectral density (PSD) of the LIGO Livingston detector in O3 \citep{2020LRR....23....3A}.
$\rho_{\text{opt}}$ is the SNR an event would have if it were ``optimally oriented'' with respect to the detector, that is, directly overhead and with its angular momentum vector pointed along the line of sight \citep{2021CQGra..38e5010C}.
In reality, GW sources have varying sky positions and angular momentum vectors.
The effect on the SNR of a source's deviation from the optimal orientation can be summarized by a multiplicative constant, $\Theta$, such that
\begin{equation}
    \rho = \rho_{\text{opt}} \Theta ,
    \label{eq:rho-true}
\end{equation}
where $\Theta$ is between zero and unity.

GW sources are typically assumed to be distributed isotropically in sky position and orientation. For a single detector, this yields a corresponding distribution for $\Theta$, described in \cite{1993PhRvD..47.2198F}.
Therefore, for each event $i$, we assign a true value $\hat{\Theta_i}$ drawn from this distribution and use it to calculate the event's true single-detector SNR $\hat{\rho}$.
The set of \emph{true} parameters for each potential event in the catalog is then $\hat{\theta}_{i} = \{\hat{\mathcal{M}_{\det}}, \hat{\eta}, \hat{\rho}, \hat{\Theta} \}_i$.

Given the true parameters, the basic procedure of generating samples from the posterior distribution of each event is to draw an observation from each event's likelihood, use that observation as the central value of the posterior distribution, and then to draw samples from that posterior, assuming a prior.

To obtain \emph{observed} parameters, $\theta^{\text{obs}}_i$, we need the likelihood, $\mathcal{L}_{\text{total}}(\theta_i^{\text{obs}}|\hat{\theta}_i)$.
We model each event's likelihood as
\begin{equation}
        \mathcal{L}_{\text{total}}(\theta_i^{\text{obs}}|\hat{\theta}_i) = \mathcal{L}_{\mathcal{M}}(\mathcal{M}_{\text{det},i}^{\text{obs}}) \mathcal{L}_{\eta}(\eta^{\text{obs}}_i) \mathcal{L}_{\Theta}(\Theta^{\text{obs}}_i) \mathcal{L}_{\rho}(\rho^{\text{obs}}_i) ,
        \label{eq:total-event-likelihood}
\end{equation}
where 
\begin{align}
\begin{aligned}
    \mathcal{L}_{\mathcal{M}}\left(\ln\left(\mathcal{M}_{\text{det},i}^{\text{obs}}\right)|\ln\left(\hat{\mathcal{M}}_{\text{det},i}\right), \rho^{\text{obs}}_i \right) &= \mathcal{N}\left(\ln{(\mathcal{M}_{\text{det},i}^{\text{obs}})}|\mu=\ln\left(\hat{\mathcal{M}}_{\text{det},i}\right), \sigma=\sigma^{\mathcal{M}}_{i}\left(\rho^{\text{obs}}_i\right)\right) \\
    \mathcal{L}_{\eta}\left(\eta^{\text{obs}}_{i}|\hat{\eta}_i,\rho^{\text{obs}}_i\right) &= \mathcal{N}\left(\eta^{\text{obs}}_{i}|\mu=\hat{\eta}_{i},\sigma=\sigma^{\eta}_{i},\left(\rho^{\text{obs}}_i\right)\right) \\
    \mathcal{L}_{\Theta}\left(\Theta_{\text{obs},i}|\hat{\Theta}_i,\rho^{\text{obs}}_i\right)  &= \mathcal{N}\left(\Theta^{\text{obs}}_{i}|\mu=\hat{\Theta}_i,\sigma=\sigma^{\Theta}_{i}\left(\rho^{\text{obs}}_i\right)\right) \\
    \mathcal{L}_{\rho}\left(\rho^{\text{obs}}_{i}|\hat{\rho}_i\right)  &= \mathcal{N}\left(\rho^{\text{obs}}_{i}|\mu=\hat{\rho}_i,\sigma=\sigma^{\rho}_{i}\right) .
    \label{eq:1d-likelihoods}
\end{aligned}
\end{align}
Here, $\mathcal{N}(\mu,\sigma)$ is the normal distribution with mean $\mu$ and standard deviation $\sigma$.

The standard deviations are determined by assuming the uncertainties on all parameters except for the SNR scale inversely with $\rho_{\text{obs}}$ \citep{2015PhRvD..91d2003V}.
In stationary, Gaussian noise, we expect the matched-filter SNR in a single detector to have unit variance \citep{2012PhRvD..85l2006A}, i.e. $\sigma^{\rho}_{i}=1$ for all $i$.
We therefore draw $\rho_{\text{obs}}$ for each event from $\mathcal{L}^{\rho}(\rho^{\text{obs}}_{i}|\rho_i)$.
This observed SNR will serve as the detection statistic that determines whether each event is observable.
We assume events that pass an SNR threshold of $\rho_{\text{obs},i} > 8$ in a single detector are detected.
In this way, we allow for events near threshold to fluctuate above or below threshold, emulating the actual noise process in the detectors.
Of the events that make it through detection, we randomly select 69 of them to constitute a mock catalog with the same number of BBHs as were analyzed by \cite{O3b_pop}.
The standard deviations for $\mathcal{M}_{\text{det}}$, $\eta$, and $\Theta$ of the detected events are calculated via
\begin{align}
\begin{aligned}
    \sigma^{\mathcal{M}}_{i}(\rho^{\text{obs}}_i) &= u_{\mathcal{M}} / \rho^{\text{obs}}_{i} \\
    \sigma^{\eta}_{i}(\rho^{\text{obs}}_i) &= u_{\eta} / \rho^{\text{obs}}_{i} \\
    \sigma^{\Theta}_{i}(\rho^{\text{obs}}_i) &= u_{\Theta} / \rho^{\text{obs}}_{i},
    \label{eq:mock-sigmas}
\end{aligned}
\end{align}
where we have chosen $u_{\mathcal{M}}=0.08\Msun$, $u_{\eta}=0.022$, and $u_{\Theta} = 0.21$ to match uncertainties in these parameters typical of events observed in O3.

Observed values for all parameters are drawn from Equation~\ref{eq:total-event-likelihood} with standard deviations defined in Equation~\ref{eq:mock-sigmas}.
With $\theta_i^{\text{obs}}$ in hand, we are now ready to construct a posterior distribution.
We apply the following priors:
\begin{align}
    \begin{aligned}
        \pi(\mathcal{M}_{\text{det}}) &= U(0\Msun,500\Msun) \\
        \pi(\eta) &= U(0,0.25) \\
        \pi(\Theta) &= U(0,1) \\
        \pi(\rho) &= U(0,300) ,
    \end{aligned}
    \label{eq:event-priors}
\end{align}
where $U(x_1,x_2)$ is the uniform distribution with lower bound $x_1$ and upper bound $x_2$.
The bounds on $\eta$ and $\Theta$ are chosen because those parameters are only physically defined in the domains $(0,0.25]$ and $[0,1]$, respectively.
Neither $\mathcal{M}$ nor $\rho$ are defined below zero, but the upper bounds were chosen somewhat arbitrarily: they must only be large enough that the likelihood has minimal support above them.
The posterior distributions for each parameter are then Gaussians centered on the observed value, with standard deviations defined in Equation~\ref{eq:mock-sigmas}.
They are therefore the same as the distributions in Equation~\ref{eq:1d-likelihoods}, but with the role of the true and observed values switched.
We then simulate multiple-dimensional posterior samples for each event by drawing \result{5000} independent samples\footnote[5]{We use \result{5000} samples to optimize the speed of population inference while also ensuring the number of effective samples used for Monte Carlo sums in the population inference always satisfies the criterion outlined in~\cite{farr_accuracy_2019}.
That criterion has since been shown to be insufficient and has been superseded by~\cite{essick_precision_2022}, but we utilize the former for consistency with the analysis performed in \cite{O3b_pop}.
However, users of the \texttt{GWMockCat} package can easily modify the number of posterior samples to suit their needs.}
of detector-frame chirp mass, symmetric mass ratio, and $\Theta$ from the posterior.
Explicitly,
\begin{align}
\begin{aligned}
    \log{\mathcal{M}_{\text{det},i}} &\sim  \mathcal{N}(\mu=\ln{\left(\mathcal{M}^{\text{obs}}_{\text{det},i}\right), \sigma=\sigma^{\mathcal{M}}_{i})} \\
    \eta_{i} &\sim \mathcal{N}(\mu=\eta^{\text{obs}}_i,\sigma=\sigma^{\eta}_{i}) \\
    \Theta_{i} &\sim \mathcal{N}(\mu=\Theta^{\text{obs}}_i,\sigma=\sigma^{\Theta}_{i}) \\
    \rho_{i} &\sim \mathcal{N}(\mu=\rho^{\text{obs}}_{i},\sigma=\sigma^{\rho}_{i})
    .
    \label{eq:observed-parameters}
\end{aligned}
\end{align}

Realistic correlations between other parameters such as component masses and redshift are obtained by transforming samples in $\{\mathcal{M}_{\text{det}}, \eta, \Theta, \rho\}$--space to $\{m_1, m_2, z\}$--space.
When necessary, we convert between luminosity distance and redshift using the cosmological parameters presented in \cite{planck_collaboration_planck_2016} so as to maintain consistency with the conventions used in \citep{gwtc1,gwtc2,gwtc3}.

The induced prior on $m_1, m_2$, and $z$ is therefore not uniform in those parameters.
This is reasonable, so long as users appropriately transform the prior when doing population inference on source-frame component masses and redshift.
We therefore provide a module in \texttt{GWMockCat} that performs these transformations.
For the case of this analysis, we opt to re-weigh the samples to a prior that is uniform in detector-frame component mass and proportional to the square of the luminosity distance in order to mimic the priors used in the standard LVK analysis \citep{gwtc1,gwtc2,gwtc3}.

The fact that Equation~\ref{eq:total-event-likelihood} is separable up to dependence on $\rho_{\text{obs},i}$ means that once $\rho_{\text{obs},i}$ is calculated for a given event, samples for $\mathcal{M}_{\text{det}}, \eta_{\text{obs}}, \Theta_{\text{obs}}$, and $\rho_{\text{obs}}$ can be drawn independently from each other.
This approximate independence is due, in part, to the fact that detector-frame chirp mass, symmetric mass ratio, SNR, and $\Theta$ are the best-measured parameters of any compact binary coalescence signal.
This fact saves considerable computational resources, allowing for many mock event posteriors to be generated quickly on a single CPU\footnote[6]{For example, a catalog of $100$ events can be generated in \result{$\mathcal{O}(10)$ seconds.}}.


We generate sensitivity estimates along with our mock catalogs to ensure that the selection function is calculated consistently to the event selection criteria \citep{essick_consistency_2022}.
To do this, we draw \result{$2 \times 10^7$} independent samples in $m_1, m_2, z$, and $\Theta$ from the following distribution:
\begin{equation}
    p(m_1,m_2,z,\Theta) \propto m_1^\alpha \left(\frac{m_2}{m_1}\right)^\beta \frac{dV_c}{dz}(1+z)^{\kappa-1} p(\Theta),
    \label{eq:pdraw}
\end{equation}
where we have chosen $\alpha = 2.35$, $\beta =1.70$, and $\kappa = 2.7$, and $p(\Theta)$ is the distribution described in \cite{1993PhRvD..47.2198F} which corresponds to isotropically oriented sources that are also isotropically positioned on the sky.
We truncate this distribution below $m_2 = 1\Msun$, above $m_1 = 200\Msun$, and above $z=4$, and confirm that there are no mock posterior samples outside of those ranges.
We will refer to these draws as ``injections.''
We then calculate an optimally-oriented SNR for each injection using the same PSD as was used for the mock observations, and compute a true SNR using Equation~\ref{eq:rho-true}.
We emulate noise fluctuations in SNR in the same way we do for mock observations, namely by using Equation~\ref{eq:1d-likelihoods}, so that each injection has a corresponding observed SNR.
Injections can then be subject to the same selection criteria as our mock observations when performing a population inference (in our case, $\rho_{\text{obs}} > 8$).

We validate this process by constructing a mock catalog from a known distribution with fixed hyperparameters, and then fitting the same distribution to our mock catalog, but allowing the hyperparameters to vary.
We then verify that the recovered hyperparameters are consistent with those used to generate the mock catalog.
The result is shown in Appendix~\ref{appendix:MDC}, along with additional validation studies.

\section{Validation of Mock Catalogs}
\label{appendix:MDC}

In this Appendix, we validate the process of creating mock event posteriors and catalogs from a known underlying population outlined in Appendix~\ref{appendix:mockPE}.
For this process, we use the same simulated catalogs utilized in Section~\ref{sec:results-2}. The simulated underlying population is described by $p_{\text{mock}}(m_1,m_2,z|\Lambda_{\text{mock}})$, where $\Lambda_{\text{mock}}$ is the set of hyperparameters $\{ \alpha, \delta, m_{\text{min}}, m_{\text{max}},\beta, \kappa \}$, 
\begin{equation}
    p_{\text{mock}}(m_1,m_2,z|\Lambda_{\text{mock}}) \propto p(m_1|\alpha,\delta,m_{\text{min}},m_{\text{max}}) p(m_2|m_1,\beta) 
    p(z|\kappa),
    \label{eq:general-population-model}
\end{equation}
and the individual mass and redshift distributions are given by the following:
\begin{align}
    p(m_1|\alpha,\delta, m_{\text{min}}, m_{\text{max}}) &\propto \begin{cases}
        0 &\text{ if } m< m_{\text{min}} \\
        m_1^{-\alpha}\frac{1}{1+ f(m - m_{\text{min}}, \delta)} &\text{ if } m_{\text{min}} \leq m < m_{\text{min}}+\delta \\
        m_1^{-\alpha} &\text{ if } m\geq m_{\text{min}} + \delta \\
        0 &\text{ if }m> m_{\text{max}}
    \end{cases}
    , \\
    p(m_2|m_1, \beta) &\propto \left(\frac{m_2}{m_1}\right)^\beta ,\\
    p(z|\kappa) &\propto \begin{cases}
        0 &\text{ if } \left(z< 0 \right) \cup \left(z> z_{\text{max}}\right) \\
        \frac{dV_c}{dz}(1+z)^{\kappa-1} &\text{ otherwise }
    \end{cases}
    .
    \label{eq:smoothed-plaw-population-model}
\end{align}
This is equivalent to the \textsc{Power Law + Peak} model in~\cite{O3b_pop} and \cite{O3a_pop}, with $\lambda_{\text{peak}}$ set to 0.
We will call the population model described by Equations~\ref{eq:general-population-model}--\ref{eq:smoothed-plaw-population-model} \textsc{Smoothed Power Law}.
We generate catalogs from the model that results from setting $\Lambda_{\text{mock}}$ to the values provided in Table~\ref{tab:smoothed-plaw-hyperparams}.
These values were chosen by fitting this population model to GWTC-3 (grey band in Figure~\ref{fig:dR_dm1}) and obtaining the median \emph{a posteriori} value for each hyperparamter.

\begin{table}[h]
    \centering
    \caption{Hyperparameter values for the underlying population of mock catalogs described by \textsc{Smoothed Power Law} (Equations~\ref{eq:general-population-model}--\ref{eq:smoothed-plaw-population-model}).}
    \label{tab:smoothed-plaw-hyperparams}
    \begin{tabular}{ c c c c }
        \hline \hline  
        \textbf{Parameter} & \textbf{Description} & & \textbf{Value} \\
        \hline
        $\beta$ & Spectral index for the power law of the mass ratio distribution. &  & 1.70 \\
        $\alpha$ & Negative spectral index for the power law of the primary mass distribution. &  & 3.14 \\
        $m_{\text{min}}$ & Minimum mass of the primary mass distribution. &  & $4.56\Msun$ \\
        $m_{\text{max}}$ &  Maximum mass of the primary mass distribution. &  & $81.08\Msun$ \\
        $\delta$ & Range of mass tapering at the lower end of the mass distribution.  &  & $5.96\Msun$ \\
        $\kappa$ & Spectral index for the power law factor of the redshift distribution. & & 2.7 \\
        \hline\hline
    \end{tabular}
\end{table}

\begin{figure*}
    \centering
    \includegraphics[width=\textwidth]{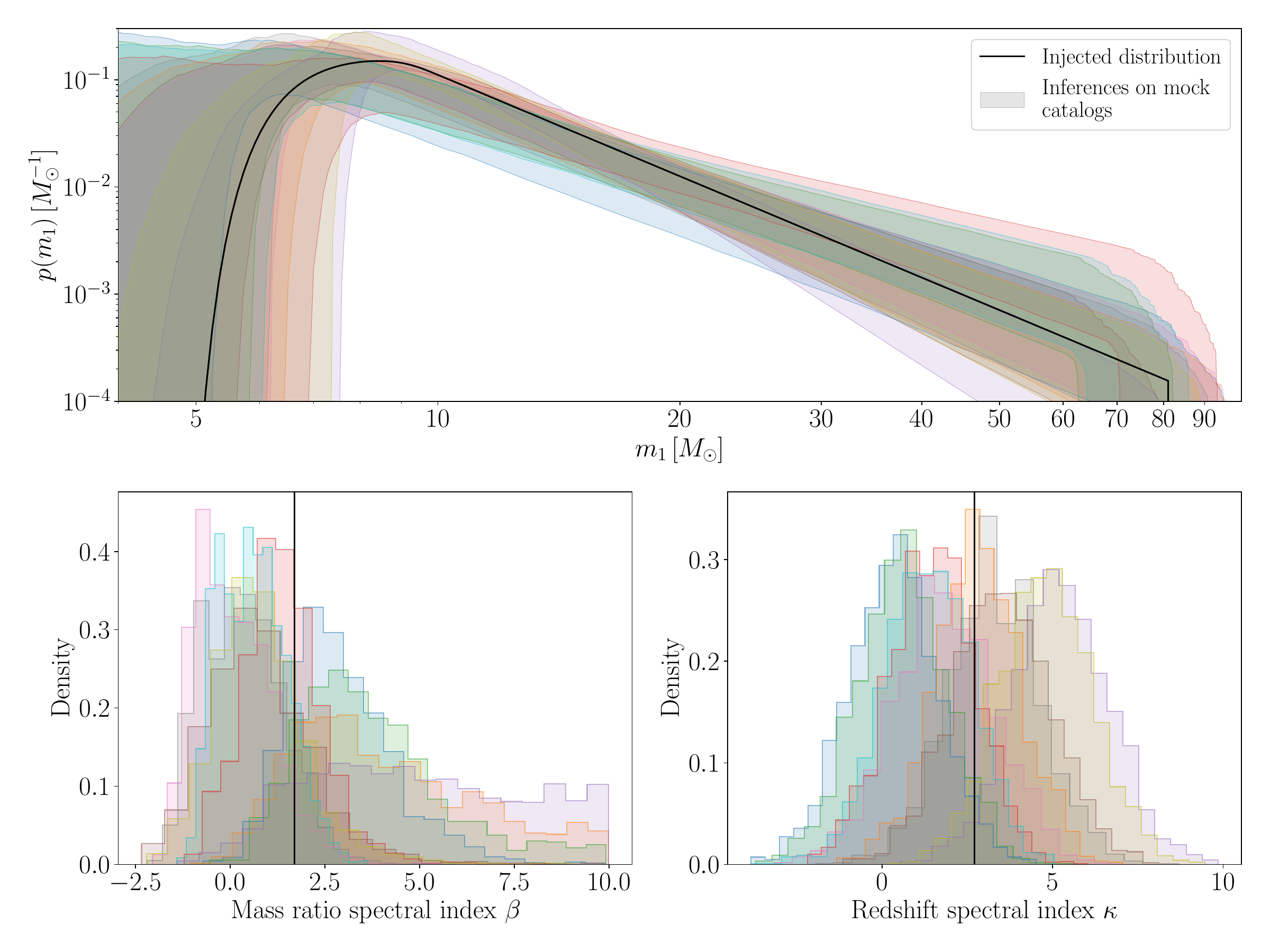}
    \caption{Injected (\emph{Solid black line}) and recovered (\emph{colored shaded bands}) distributions for \result{10} mock catalogs.
    \emph{Top:} probability density function of primary masses.
    \emph{Bottom left:} hyperposterior distribution for $\beta$, the power law spectral index of the mass ratio distribution.
    \emph{Bottom Right:} hyperposterior distribution for $\kappa$, the spectral index of the power law factor in the redshift distribution.}
    \label{fig:MDC_10_catalogs}
\end{figure*}

We validate the mock catalogs' generation by fitting them with \textsc{Smoothed Power Law} and allowing the hyperparameters to be inferred from the mock data.
We then determine whether the inferred values of the hyperparameters are consistent with the values in Table~\ref{tab:smoothed-plaw-hyperparams}.
We fit \result{100} mock catalogs of 69 events each, \result{10} results of which are shown in Figure~\ref{fig:MDC_10_catalogs}.
While there is noticeable scatter about the injected value, it is generally consistent with the recovered mass distributions: the hyperparameters of the underlying mass distribution fall within the inferred mass hyperparameters' 90\% credible intervals \result{$89.6\%$} of the time.
We therefore conclude that any biases that the mock posterior generation process introduces in the mass distribution are subdominant to the statistical uncertainties of the fit.

\begin{figure}
    \centering
    \includegraphics[width=\textwidth]{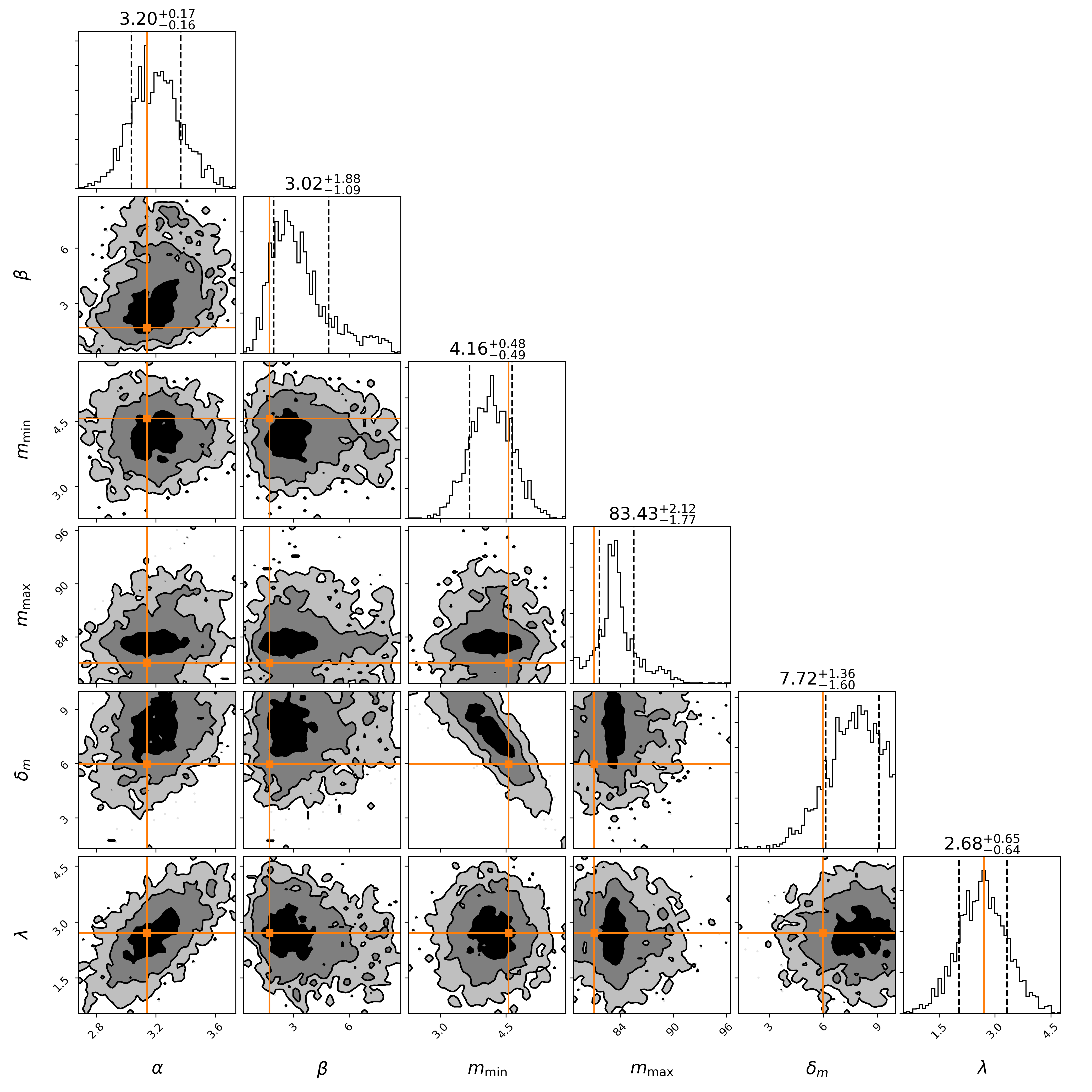}
    \caption{A corner plot of the inferred hyperposterior from a fit to a mock catalog with 345 events. 
    The injected values are shown in orange.
    The recovered hyperposterior is consistent with the injected population.
    }
    \label{fig:MDC_one_big_catalog}
\end{figure}

To further explore systematic differences caused by mock catalog generation that may be subdominant to the considerable statistical uncertainties resulting from a fit to only 69 events, we fit \textsc{Smoothed Power Law} to a single catalog of that is \result{five times larger}.
The result is shown in Figure~\ref{fig:MDC_one_big_catalog}. 
\comment{We find that the hyperparameters of the underlying distribution are consistent with the inferred hyperposterior to \result{X\%}.}
The hyperparameters of the underlying distribution seem to be consistent with the inferred hyperposterior, so we conclude that our mock event posterior generation process produces biases subdominant to measurement uncertainty typical of 345-event catalogs.
We therefore find this method of generating mock catalogs sufficient to test the significance of features identified in the mass distribution of GWTC-3.

\section{Accuracy of Mock Catalogs when Used in a Population Analysis}
\label{appendix:mockPE_accuracy}
\begin{figure}
    \centering
    \includegraphics[width=\textwidth]{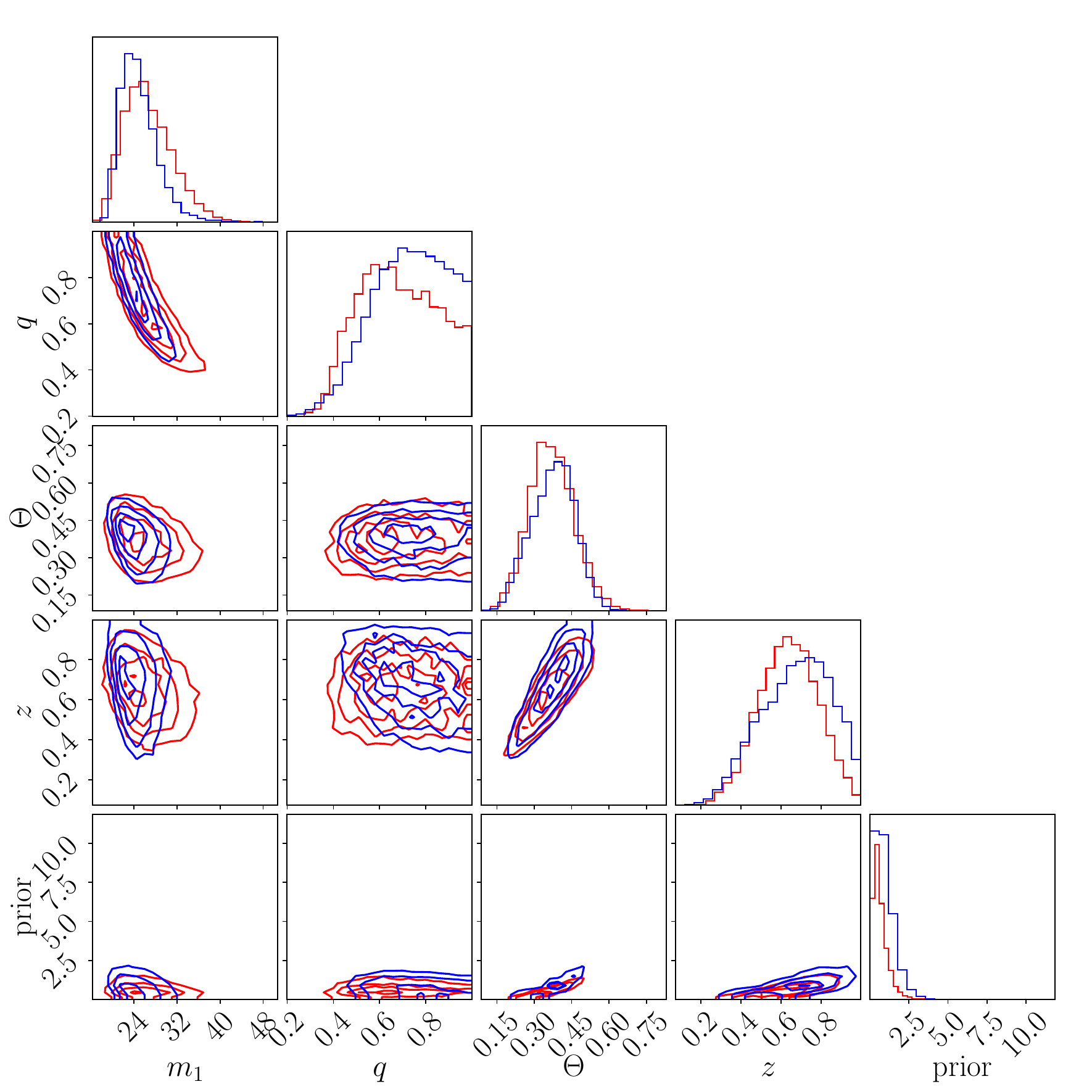}
    \caption{\new{Mock posteriors simulated with \texttt{GWMockCat} (\emph{red}) compared to posteriors made with full parameter estimation as released by the LVK (\emph{blue}) for the event GW191215\_223052. 
    The two sets of posterior samples appear consistent to the level needed for a population analysis.}}
    \label{fig:mock_PE_191215}
\end{figure}

\begin{figure}
    \centering
    \includegraphics[width=.7\textwidth]{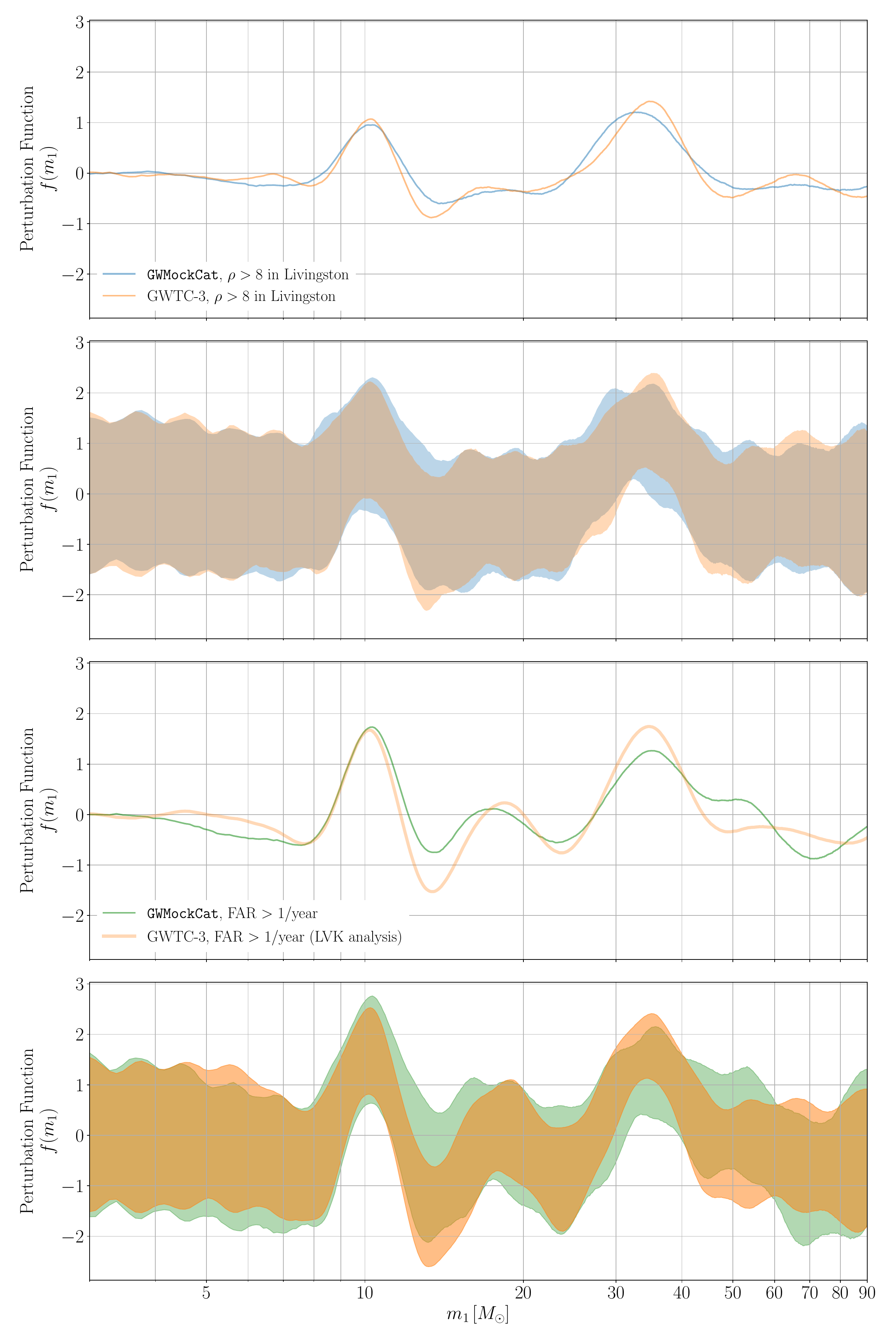}
    \caption{\new{Perturbation functions of a population fit to GWTC-3 when using LVK-released posterior samples for each event (\emph{orange}), and when using posterior samples simulated by \texttt{GWMockCat} (\emph{blue and green}).
    The two sets of figures correspond to two selection criteria, one which is analogous to the one used for mock catalogs in this work (SNR $> 8$, \emph{top two panels}), and one which is identical to that used for the LVK analysis (FAR$>1$, \emph{bottom two panels}).
    The main difference between the selection criteria is that they result in catalog sizes that are different by nearly a factor of two, and therefore the statistical error on the perturbation function is noticeably different between them.
    Using the same number of events in the simulated catalog as the LVK-released catalog results in a perturbation function that is similar in amplitude to the LVK-released population analysis.
    We conclude that the prescriptions used in \texttt{GWMockCat} are sufficient for the purposes of population analyses.
    We use the green curves in the bottom two panels for the analysis presented in the body of this paper, and the orange curves in the bottom two panels for the analysis presented in this Appendix.
    }}
    \label{fig:mock_PE_accuracy}
\end{figure}

\new{
As a second check of \texttt{GWMockCat}'s ability to simulate catalogs accurately enough to be used in a population analysis of the mass distribution of BBHs, we recreate GWTC-3 with \texttt{GWMockCat}. 
We then compare \PS{}'s fit to this mock catalog with its fit to the posterior samples released by the LVK for GWTC-3 \citep{gwtc3_data_release}.
We find that the two resulting mass distributions are consistent, and therefore conclude that the approximate prescriptions used in \texttt{GWMockCat} are sufficient to probe the mass distribution of BBHs, at least for current GW detector sensitivities.}

\new{
To simulate GWTC-3, we re-weight all GWTC-3 events to the same prior as used for sampling in \texttt{GWMockCat}, namely uniform in detector-frame chirp mass, uniform in symmetric mass ratio, and uniform in sky angle, as defined in Equation~\ref{eq:event-priors}. 
We then take the mean of the re-weighted detector frame chirp mass, symmetric mass ratio, sky angle, and single-detector SNR posteriors as the observed parameters $\theta_i^{\text{obs}}$ for each event.
Using the priors defined in Equation~\ref{eq:event-priors} and likelihoods defined in Equations ~\ref{eq:1d-likelihoods} and ~\ref{eq:mock-sigmas}, we construct mock posterior distributions on detector frame chirp mass, symmetric mass ratio, sky angle, and single-detector SNR. 
We then convert these parameters to source-frame component masses and redshift.
Finally, we re-weight back to the standard prior used in LVK's parameter estimation process.
The end result of this is shown in Figure~\ref{fig:mock_PE_191215} for GW191215\_223052, an event that was chosen at random from the catalog and happened to have a primary mass and redshift near the mode of the detected events.
The mock posteriors appear consistent with the true posteriors, and degeneracies between parameters seem to be suitably captured.
This behavior is qualitatively similar for all simulated events, though some had mock posteriors that were slightly more consistent with the full parameter estimation posterior samples, and some had mock posteriors that were slightly less consistent.}

\new{
Any population analysis must define criteria for inclusion in the population.
We apply two different criteria and report the results of both in Figure~\ref{fig:mock_PE_accuracy}.
The first, a cut at $\mathrm{SNR} > 8$ in Livingston, is chosen to be analogous to the detection criteria of the other mock catalogs, which use a single-detector SNR cut since it is not possible to run all of the pipelines necessary to produce a FAR on mock data.
The second criterion, a false alarm rate (FAR) cut at 1/yr was used to be consistent with the analysis done by the LVK on GWTC-3 \citep{O3b_pop}.
Either choice is reasonable because the selection function is known with respect to both SNR and FAR.
We can therefore use the publicly-released sensitivity estimates~\citep{O3_injections_data_release} to reconstruct the underlying, or astrophysical distribution of BBHs from either of these catalogs.
All of the metrics of feature significance presented in Section~\ref{sec:full} make use of this astrophysical distribution when comparing GWTC-3's population fit to that of mock catalogs.} 

\new{
The first criterion (single-detector SNR cut) produced a final catalog with many fewer events than the second criterion (FAR cut), with the former resulting in 36 events and the latter resulting in 69 events.
Therefore, the population analysis performed on the catalog selected by SNR has much wider hyper-posteriors than the one performed on the catalog selected by FAR.
However, \emph{these two catalogs do not appear to be systematically biased with respect to one another, nor are they systematically biased with respect to the LVK-released analysis}.
This is again because the selection function is known with respect to both of these criteria, and therefore the reconstructed astrophysical distributions are consistent.}

\new{
Interestingly, the mock catalog with only 36 events in it still finds the peak at $\sim35\Msun$ to be significant, with the perturbation function excluding zero to $<5\%$.
However, other features are not well enough resolved to appear significant with only 36 events.
We find very little difference between the full parameter estimation perturbation function fit (orange bands in Figure~\ref{fig:mock_PE_accuracy}) and the mock catalog (blue band) in the case of a single-detector SNR cut (top two panels).
We take this to mean that for high-SNR events, the mock parameter estimation is sufficient for use in population analyses of the mass distribution, at least for O3-like detector sensitivities.
}

\new{
Using the same events in the simulated catalog as the LVK-released catalog results in a perturbation function that is similar in amplitude to the LVK-released analysis (lower two panels of Figure~\ref{fig:mock_PE_accuracy}).
However, the width of the perturbation function's hyperposterior is slightly inflated in the mock catalog case.
This may be because the mock parameter estimation scales event posterior widths inversely with SNR, so mock posterior widths are over-estimated with respect to full parameter estimation for the events that meet the FAR threshold but have low SNR.}

\new{In order to be as consistent as possible in our comparisons of mock catalogs to GWTC-3, we use the perturbation function obtained by analyzing the \texttt{GWMockCat} version of GWTC-3 for all comparisons to mock catalogs in Section~\ref{sec:full}. 
In this Appendix, we repeat the analysis performed in Section~\ref{sec:full}, but instead used the perturbation function released by the LVK in \cite{O3b_pop} in lieu of the perturbation function obtained by fitting \PS{} to the \texttt{GWMockCat} version of GWTC-3.
In other words, the main text uses the green curves in Figure~\ref{fig:mock_PE_accuracy}, and we repeat the analysis using the orange curves in the bottom two panels of Figure~\ref{fig:mock_PE_accuracy} in this Appendix.}

\new{We find that using the LVK-released perturbation function increases the Bayesian significance of all features relative to the \texttt{GWMockCat} reproduction.
This is consistent with the conclusions reached in Figure~\ref{fig:mock_PE_accuracy}: all hyperposteriors narrow slightly when using the LVK-released version of parameter estimation, but there is no systematic shift as a function of primary mass or any other parameter.
When using the LVK-released perturbation function, \result{None} of the \result{300} $\{g_j(f_{\max})\}$ exclude zero to the same percentile as $g_{\text{GWTC-3}}(f(35\Msun))$ or $g_{\text{GWTC-3}}(f(10\Msun))$, and \result{$1.3\%$} of the $\{g_j(f_{\max})\}$ exclude zero to the same percentile as $g_{\text{GWTC-3}}(f(14\Msun))$.
These values are smaller than those presented in Section~\ref{sec:zero-exclusion}, but lead to the same conclusions: the $10\Msun$ and $35\Msun$ peaks are difficult to reproduce with featureless catalogs, but the $14\Msun$ dip is not.
Performing full parameter estimation on mock catalogs would also likely narrow the hyperposteriors for those catalogs, in turn increasing the significance of the peaks seen in their the perturbation functions.
If this were to be the case, the fraction of mock catalogs that can reproduce features in the GWTC-3 distribution would likely be similar to those found in Section~\ref{sec:zero-exclusion}.
A reproduction of the analysis presented in Section~\ref{sec:ks_perc} with the LVK-released perturbation function also finds similar results to that done on the \texttt{GWMockCat} perturbation function. 
We therefore conclude that the results presented Section~\ref{sec:full} are robust to the procedure used for parameter estimation of GWTC-3 events.}



\end{document}